\documentclass[acmsmall,screen]{acmart}

\AtBeginDocument{%
  }

\newcommand{\scheme}{TraceLite}
\usepackage{enumitem}
\usepackage{algorithm}
\usepackage{algpseudocode}
\usepackage{booktabs}
\usepackage{tabularx}
\usepackage{array}
\usepackage{multirow}
\usepackage[table]{xcolor}

\usepackage{pifont}

\definecolor{opticalgreen}{RGB}{235,247,232}
\definecolor{overlapyellow}{RGB}{255,248,220}

\usepackage{graphicx}
\usepackage{tcolorbox}
\usepackage{subfigure}

\newcommand{\cmark}{\color{green!60!black}\ding{51}}
\newcommand{\xmark}{\color{red!60!black}\ding{55}} 
\usepackage{titletoc}

\setcopyright{acmlicensed}
\copyrightyear{2018}
\acmYear{2018}
\acmDOI{XXXXXXX.XXXXXXX}

\acmConference[Conference acronym 'XX]{Make sure to enter the correct
  conference title from your rights confirmation email}{June 03--05,
  2018}{Woodstock, NY}

\acmISBN{978-1-4503-XXXX-X/2018/06}

\begin{document}




\title{\scheme: Efficient Trace Reduction for SWE Agents via Optical Compression and Redundancy Removal}

\author{Jingyu Xiao}
\affiliation{%
  \institution{The Chinese University of Hong Kong}
  \city{Hong Kong}
  \country{China}
}
\email{jyxiao@link.cuhk.edu.hk}

\author{Qianxi Liu}
\affiliation{%
  \institution{Dalian University of Technology}
  \city{Dalian}
  \country{China}}
\email{lqx19152881917@gmail.com}

\author{Yintong Huo}
\authornote{Yintong Huo is the corresponding author.}
\affiliation{%
 \institution{Singapore Management University}
 \city{Singapore}
 \country{Singapore}}
\email{ythuo@smu.edu.sg}

\author{Yuan Jiang}
\affiliation{%
  \institution{Harbin Institute of Technology}
  \city{Harbin}
  \country{China}}
\email{jiangyuan@hit.edu.cn}

\author{Michael R. Lyu}
\affiliation{%
  \institution{The Chinese University of Hong Kong}
  \city{Hong Kong}
  \country{China}}
\email{lyu@cse.cuhk.edu.hk}

\renewcommand{\shortauthors}{Jingyu Xiao et al.}

\begin{abstract}
Software engineering agents solve repository-level tasks through iterative tool use, but their continuously growing interaction traces incur substantial token consumption and inference overhead. Existing compression approaches often require model training, auxiliary LLM calls, or assumptions about specific content such as source code. We present \scheme, a simple yet effective, training-free framework for compressing heterogeneous SWE-agent observations. Our preliminary study shows that optical rendering preserves substantial textual and semantic information under moderate compression, while agent trajectories exhibit both cross-step and phase-level redundancy. Motivated by these findings, \scheme \ preserves recent observations as high-fidelity text, renders historical observations as compact images, removes overlapping content, and discards phase-expired observation payloads after code editing. It requires neither offline training nor auxiliary LLM inference. We evaluate \scheme \ on SWE-bench Verified and DeepSWE using GPT-5.6-Luna and Gemini-3.8-Flash as backbone models. On SWE-bench Verified, \scheme \ achieves 78.25\% resolve rate, outperforming the evaluated compression baselines while reducing average input-token consumption and API cost by 33.05\% and 21.30\%, respectively. Moreover, it introduces only 0.438 seconds of compression latency per observation. These results demonstrate that \scheme \ improves SWE-agent efficiency with low processing overhead and without compromising issue-resolution capability. Our code is available at \url{https://github.com/xjywhu/TraceLite}.
\end{abstract}

\ccsdesc[500]{Software and its engineering~Automatic programming}
\ccsdesc[300]{Computing methodologies~Artificial intelligence}

\keywords{Large Language Models, Coding Agent, Token Compression}


\maketitle

\section{Introduction}

Large language models have evolved from passive code-generation tools into coding agents that autonomously interact with software development environments~\cite{yang2024sweagent,wang2024openhands}. Equipped with interfaces for repository search, file editing, command execution, and testing, these agents have been applied to tasks such as fault localization, program repair, test generation, and issue resolution. Rather than producing a solution in a single invocation, a coding agent iteratively inspects the repository, invokes tools, interprets their outputs, and adjusts its subsequent actions. SWE-agent exemplifies this paradigm by introducing an agent--computer interface through which an LLM can navigate repositories, modify files, and execute tests~\cite{yang2024sweagent}. Its evaluation on SWE-bench, a benchmark constructed from real-world GitHub issues, demonstrates the potential of agents to perform repository-level debugging and patch generation~ \cite{jimenez2024swebench}.

However, this iterative workflow causes the interaction trajectory to grow continuously~\cite{xiao2026agentdiet,gao2026more}. Tool observations such as source code, search results, command outputs, and test logs are repeatedly included in subsequent prompts. Retaining redundant, outdated, or irrelevant observations increases inference cost and context window pressure and may distract the model from evidence needed for its current decision. Effective context management is therefore essential for compressing historical interaction traces while preserving information useful for future actions.

Existing \textbf{context compression} methods differ in their targets, redundancy criteria, and context-handling strategies, as summarized in Table~\ref{tab:method_taxonomy}. LLMLingua2~\cite{pan2024llmlingua2} learns token-level importance using a dedicated compressor, whereas LongCodeZip~\cite{shi2025longcodezip} uses conditional perplexity to prune code at the function and block levels. SWE-Pruner~\cite{wang2026swepruner} instead employs a trained neural pruner to perform line-level code filtering. SeeRepo~\cite{ma2026llmagentscoderepositories} constructs and renders repository graphs to guide code exploration, augmenting the agent's context rather than compressing its accumulated trajectory. AgentDiet~\cite{xiao2026agentdiet} directly reduces agent trajectories, but relies on an auxiliary reflection LLM to identify and remove redundant or expired information. Consequently, existing approaches require either learned components, auxiliary model inference, or repository-specific preprocessing, and do not provide a training-free and auxiliary-free solution for context compressing.

\begin{table*}[ht]
    \centering
    \caption{Taxonomy of context compression methods,  OR denotes optical rendering.}
    \label{tab:method_taxonomy}
    \setlength{\tabcolsep}{1.5pt}

    \resizebox{0.99\textwidth}{!}{
    \begin{tabular}{lllll}
        \toprule
        \textbf{Method}
        & \textbf{Target}
        & \textbf{Redundancy Judge}
        & \textbf{Context Handle}
        & \textbf{Extra Cost} \\
        \midrule

        LLMLingua2~\cite{pan2024llmlingua2}
        & Single prompt
        & Token-level importance
        & Token-level pruning
        & Trained compressor \\

        LongCodeZip~\cite{shi2025longcodezip}
        & Code function
        & Perplexity-based relevance
        & Block-level pruning
        & Auxiliary LLM \\

        AgentDiet~\cite{xiao2026agentdiet}
        & Trajectory
        & LLM-as-judge
        & LLM-guided reduction
        & Additional LLM calls \\

        SeeRepo~\cite{ma2026llmagentscoderepositories}
        & Code Repo
        & --
        & Graph augmentation
        & \shortstack[l]{LLM-based graph\\
                 information extraction} \\

        SWE-Pruner~\cite{wang2026swepruner}
        & Code lines
        & Line-level relevance
        & Line-level pruning
        & Trained pruner \\

        \midrule

        \textbf{\scheme \ (Ours)}
        & \textbf{Trajectory}
        & \textbf{Overlap and Edit-trigger}
        & \textbf{OR and remove}
        & \textbf{None} \\

        \bottomrule
    \end{tabular}
    }
\end{table*}

These limitations raise a natural question: can interaction traces be compressed effectively without training a dedicated compressor or invoking an auxiliary model? Answering this question requires examining two complementary aspects of an agent trajectory: \emph{representation} and \emph{content}. First, can verbose observations be represented more compactly while preserving the information required by the agent? Second, does every observation remain necessary throughout the entire issue-resolution process?

To answer these questions, we conduct a preliminary study of SWE-agent trajectories. Regarding representation, we find that rendering observations as images preserves substantial character-level and semantic information under moderate compression. However, rendering the entire trajectory degrades issue-resolution performance because recent observations often contain precise evidence needed for the next action. Regarding content, we identify two forms of redundancy. First, observations from different interaction steps exhibit substantial textual overlap. Second, localization and diagnostic observations become largely unnecessary after the agent applies a concrete code modification, as their information has been incorporated into the updated repository state. These findings motivate a hybrid strategy that preserves recent evidence, compresses historical observations, and removes redundant content.

Motivated by these findings, we propose \scheme, a simple yet effective training-free framework for compressing SWE-agent interaction traces. \scheme\ divides the trajectory into recent and historical windows: recent observations are retained as high-fidelity text, while observations leaving the recent window are rendered as compact images. Within the recent window, \scheme\ detects textual overlap and removes duplicated content from earlier observations while preserving their tool calls and non-overlapping information. Moreover, when the agent performs a code-editing action, \scheme\ treats the edit as a phase transition and clears the preceding observation payloads while keeping the historical tool-call sequence. The complete procedure requires neither model training nor auxiliary LLM calls and supports heterogeneous tool observations.

Our contributions are summarized as follows:

\begin{itemize}[leftmargin=*, nosep]

    \item We conduct a preliminary study of SWE-agent trajectories and
    identify three opportunities for compression: optical rendering
    preserves substantial information under moderate compression,
    observations across different steps contain considerable textual
    overlap, and pre-edit observations become largely unnecessary after
    the repository is modified.

    \item We propose \scheme, a simple yet effective trace-compression
    framework that combines historical optical
    compression and redundancy removal. It requires neither model training nor auxiliary LLM calls and supports heterogeneous tool observations.
    
    \item We extensively evaluate \scheme \ on SWE-bench Verified and DeepSWE using GPT-5.6-Luna and Gemini-3.8-Flash. On SWE-bench Verified, \scheme \ achieves the highest resolved rate of 78.25\%, while reducing input tokens and API cost by 33.05\% and 21.30\%, respectively. It requires no training or auxiliary LLM calls and adds only 0.438 seconds of latency per observation.

\end{itemize}


\section{Preliminary Study}
\label{sec:preliminary}


Before presenting the design of \scheme, we conduct a preliminary study to examine two properties of SWE-agent interaction traces. First, we investigate whether historical observations can be represented more compactly without losing their textual details or task-relevant semantics. Second, we examine whether observations produced at different interaction steps contain redundant or overlapping information. These two questions motivate the optical compression and trajectory refinement components of \scheme, respectively.  We collect interaction trajectories generated by GPT-5.6-Luna~\cite{openai2026gpt56luna} using the mini-SWE-agent~\cite{yang2024sweagent} framework on SWE-bench Verified~\cite{jimenez2024swebench}. 



\subsection{RQ1: Can Observations Rendered as Images Preserve Information?}
\label{sec:preliminary_rq1}

Recent OCR-based context compression methods~\cite{wei2025deepseekocr,shi2026codeocr} demonstrate that visually rendered text can be represented using fewer tokens than its original textual form. Inspired by this observation, we consider rendering historical SWE-agent observations as images. However, software engineering observations require fidelity at both the character and semantic levels. At the character level, the agent must accurately recognize file paths, identifiers, line numbers, and error messages. At the semantic level, it must understand information such as program behavior, the purpose of a code fragment, and the cause of a test failure. Even a minor recognition error or semantic distortion may mislead subsequent agent decisions. We therefore investigate whether optical rendering preserves both the textual details and task-relevant semantics of agent observations.




We randomly sample 126 observations covering the different outputs encountered during issue resolution like source code, file paths, search results, and execution logs. Following CodeOCR~\cite{shi2026codeocr}, we control the compression ratio by adjusting the resolution of the rendered observation. An MLLM divides an image of resolution $W\times H$ into fixed-size patches of size $p$, resulting in approximately $(W/p)\times(H/p)$ visual tokens. We define $k\times$ compression such that the visual token count is $1/k$ of the original text token count. Each observation is first rendered with fixed font and layout settings at a base resolution of $2240\times2240$ and then resized using bilinear downsampling.

(1) \textbf{Text reconstruction.} We first evaluate whether the textual content of an observation can be recovered from its rendered image. Specifically, GPT-5.6-Luna is given only the rendered observation image and instructed to reconstruct the original text. We evaluate reconstruction quality by comparing the reconstructed text $\hat{O}$ with the ground-truth observation $O$ using Levenshtein edit distance. Because the raw distance is affected by sequence length, we normalize it to define the character-level reconstruction similarity as
$\operatorname{Sim}(O,\hat{O})
= 1 - D_{\mathrm{lev}}(O,\hat{O})
  / \max\left(|O|,|\hat{O}|\right)$,
where $D_{\mathrm{lev}}(O,\hat{O})$ denotes the character-level Levenshtein edit distance between $O$ and $\hat{O}$. The resulting score ranges from 0 to 1, with higher values indicating greater fidelity to the original observation.

(2) \textbf{Semantic question answering.}
Exact text reconstruction provides a strict measure of character-level preservation, but minor transcription differences do not necessarily imply the loss of task-relevant semantics. We therefore apply the following three fixed, observation-independent questions to every observation: (1)~\emph{What type of software-engineering information does this observation contain, and to which functional category does it belong?} (2)~\emph{What is the primary purpose of this observation in the issue-resolution process, including the object it concerns and the intended outcome?} and (3)~\emph{What are the main operations, results, state changes, or errors reported by this observation?} The first question identifies the broad functional category, the second
captures the observation's role and objective, and the third examines its core operational semantics.

\begin{figure}[h]
    \centering
    \begin{minipage}{0.48\linewidth}
        \centering
        \subfigure[Text Reconstruction.]{
            \label{subfig:text_reconstruction}
            \includegraphics[width=0.46\linewidth]
            {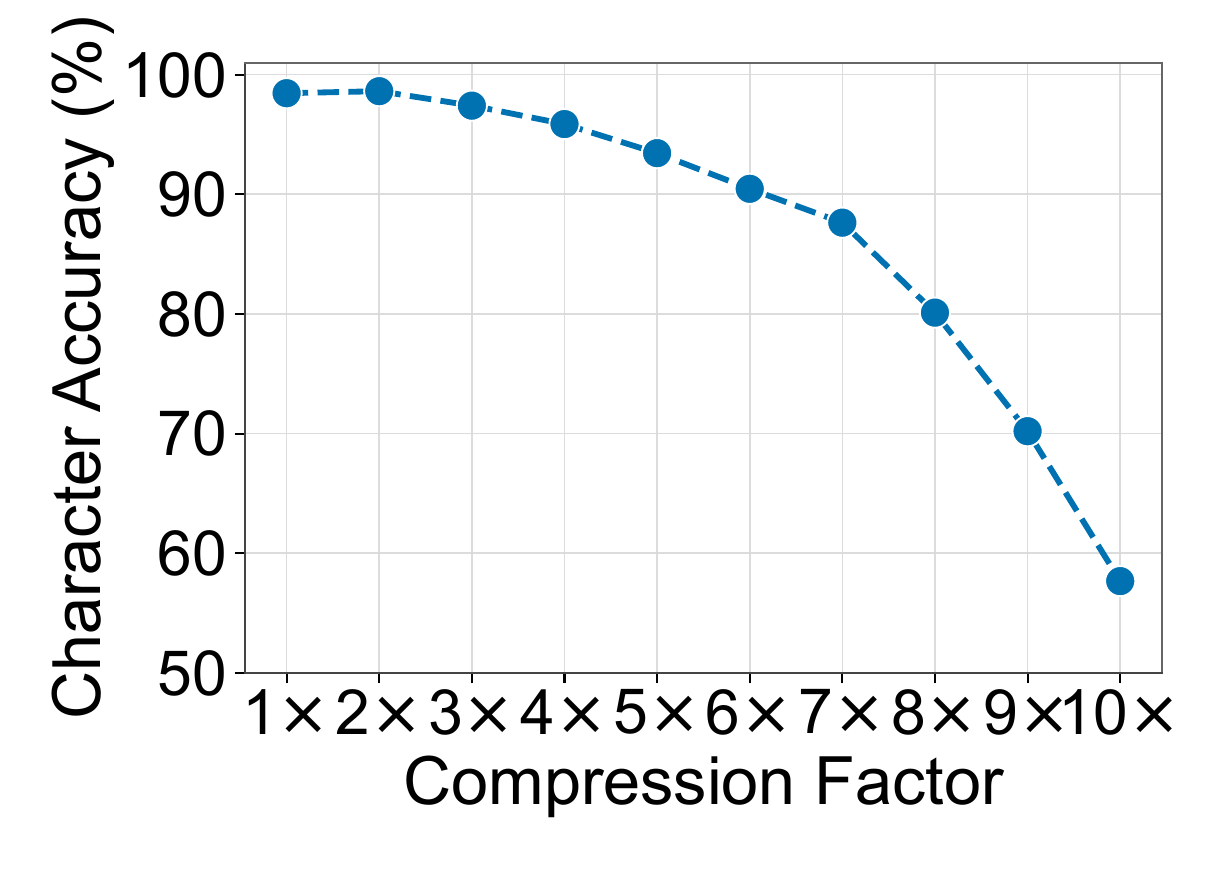}
        }
        \subfigure[Semantic QA.]{
            \label{subfig:semantic_qa}
            \includegraphics[width=0.46\linewidth]
            {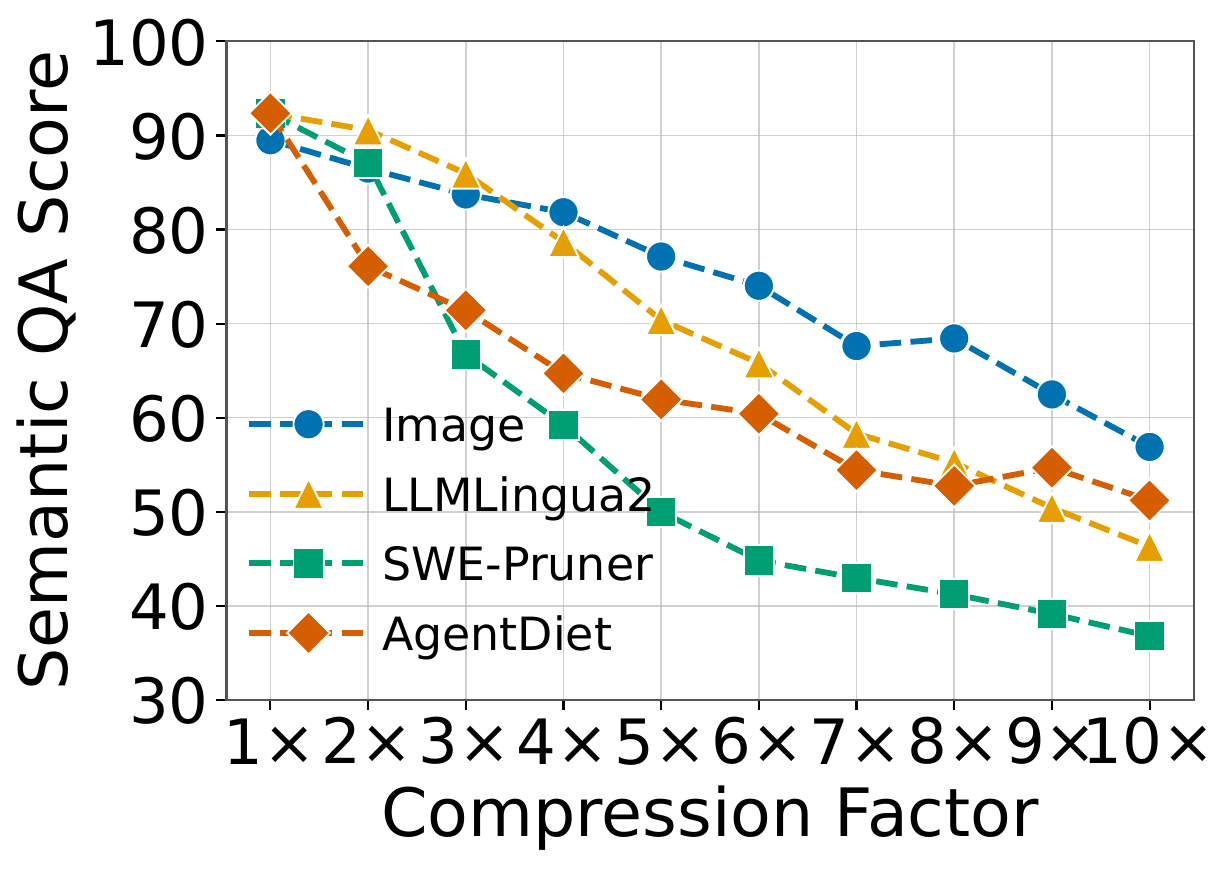}
        }
        \caption{OCR compression performance.}
        \label{fig:observation_ocr}
    \end{minipage}
    \hfill
    \begin{minipage}{0.48\linewidth}
        \centering
        \subfigure[Overlap Ratio.]{
            \label{subfig:overlap_ratio}
            \includegraphics[width=0.45\linewidth]
            {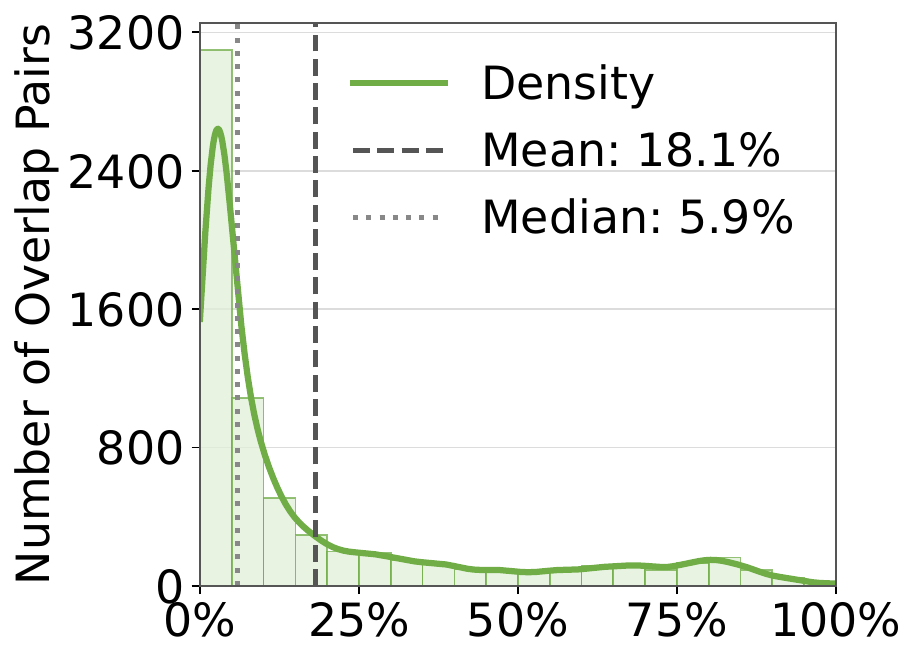}
        }
        \subfigure[Overlap Step Gap.]{
            \label{subfig:overlap_step}
            \includegraphics[width=0.45\linewidth]
{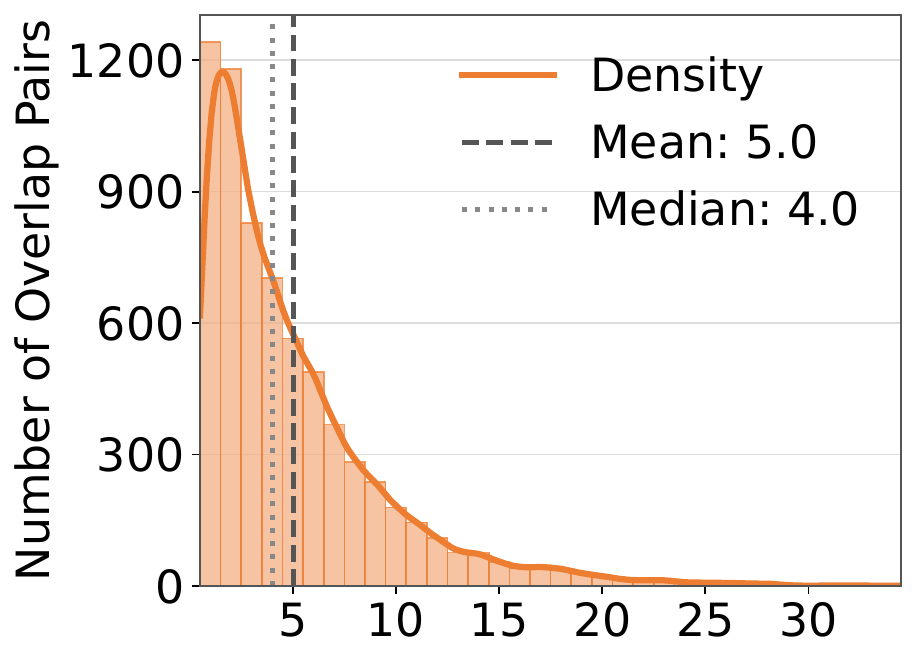}
        }
        \caption{Inter-observations redundancy.}
        \label{fig:observation_overlap}
    \end{minipage}
\end{figure}

In the \emph{text setting}, GPT-5.6-Luna answers these questions using the
original textual observation. In the \emph{image setting}, it answers the
same questions using only the rendered image. Both settings use identical
instructions and decoding configurations. GPT-6-Sol is used to compare the text-conditioned answer as the reference with the image-conditioned answer as the candidate. The judge is not informed of the corresponding input modalities and evaluates only macro-level semantic preservation. It independently assigns a score from 0 to 100 for \emph{category}, \emph{purpose}, and \emph{operations}, where higher scores indicate stronger functional equivalence, we finally report the normalized score. We repeat each judgment 3 times and use the mean score. As shown in Fig.~\ref{fig:observation_ocr}, optical rendering
preserves both textual and semantic information well under moderate
compression. Fig.~\ref{subfig:text_reconstruction} shows that
character accuracy remains above 95\% through $4\times$ compression and
above 90\% through $6\times$, but declines sharply beyond $7\times$,
reaching 57.67\% at $10\times$. For semantic preservation,
Fig.~\ref{subfig:semantic_qa} shows that the image-based representation
degrades more slowly than the compression baselines and achieves the
highest QA score from $4\times$ onward. Even at $10\times$, it retains a
semantic QA score of approximately 57, compared with 51 for AgentDiet,
46 for LLMLingua2, and 37 for SWE-Pruner. These results demonstrate that
observation-as-image representations offer strong information
preservation under moderate compression and remain more robust than
text-based compression methods at higher compression factors.

\begin{tcolorbox}[
    colback=gray!20,
    colframe=gray!20,
    width=\columnwidth,
    left=0.05in,
    right=0.05in,
    top=0.05in,
    bottom=0.05in
]
\textbf{Observation 1:} Optical rendering maintains high character
fidelity under moderate compression and preserves semantic information
more robustly than text-based compression methods, although aggressive
compression degrades both.
\end{tcolorbox}

\subsection{RQ2: Redundancy across Agent Observations}
\label{sec:preliminary_rq2}

We investigate two complementary forms of redundancy in agent trajectories. The first is \emph{content redundancy}, where observations from different steps contain overlapping textual information. The second is \emph{phase-level redundancy}, where observations useful during localization and diagnosis become unnecessary after the agent transitions to code editing. We analyze these two forms separately. We analyze trajectories generated by GPT-5.6-Luna using mini-SWE-agent on SWE-bench Verified and group the corresponding tool calls into four categories: localization, code editing, testing, and patch review.


\textbf{Cross-Step Content Redundancy.}
We measure content redundancy at the line level. Specifically, we split each observation $O_i$ into
non-empty lines and represent it as a set of normalized lines $S(O_i)$.
We then calculate the overlap ratio between two observations using the
Jaccard coefficient: $\operatorname{Overlap}(O_i,O_j) =
\frac{|S(O_i)\cap S(O_j)|}
{|S(O_i)\cup S(O_j)|}.$
A higher value indicates that the two observations share a larger proportion
of identical lines. For each observation pair with non-zero line-level
overlap, $(O_i,O_j)$, we additionally record its step gap
$\Delta_{i,j}=j-i$ to measure how long duplicated content persists in the
trajectory.

As shown in Fig.~\ref{subfig:overlap_ratio}, observation pairs exhibit a mean overlap ratio of 18.1\% and a median of 5.9\%. The distribution spans a wide range, with a notable proportion of pairs exhibiting substantial redundancy and some sharing more than 75\% of their content. Overall, 41.75\% of observations overlap with at least one other observation in the same trajectory, indicating that cross-step content redundancy is widespread in full-history agent trajectories. Fig.~\ref{subfig:overlap_step} shows that overlapping observations are typically close to each other, with a mean step gap of 5 and a median of 4. Although a small number of overlapping pairs are separated by more than 30 steps, most redundancy occurs among recent interactions, motivating overlap removal within the recent window.



\begin{figure}[ht]
\centering
\includegraphics[width = .99\textwidth]{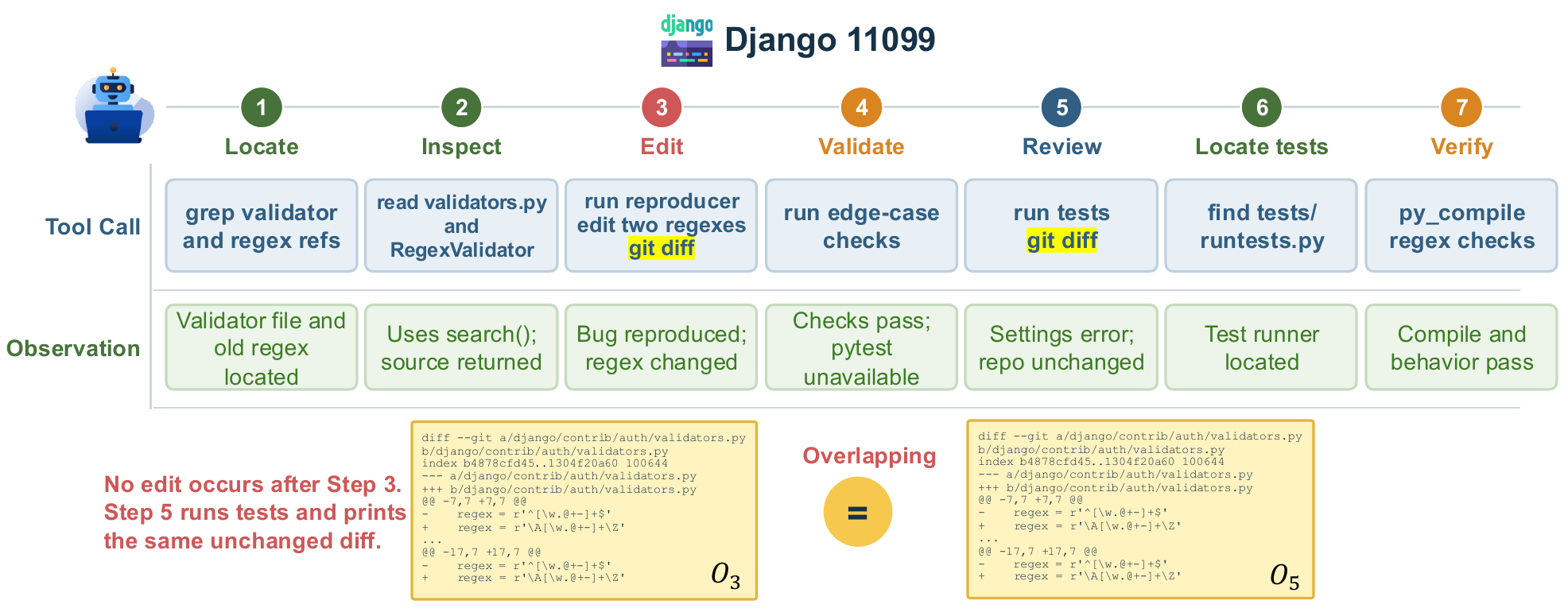}
\caption{An example of observation overlap in a Django issue-resolution trajectory. Steps~3 and~5 return the same Git diff despite performing different tool operations.}
\label{fig:overlap_example}
\end{figure}

Fig.~\ref{fig:overlap_example} presents a representative trajectory from Django issue 11099. At Step~3, the agent reproduces the bug, edits two regular expressions, and receives the resulting Git diff. The agent then validates the modification at Step~4 without changing the repository. At Step~5, it attempts to run the tests and invokes \texttt{git diff} again. Although the test command produces a new configuration error, the observation repeats the same diff previously returned at Step~3 because no intervening edit has occurred. This case illustrates that different tool calls can produce observations containing the same repository state. Retaining both complete observations therefore duplicates the diff.

\begin{figure*}[ht]
    \centering

    \begin{minipage}[t]{0.17\textwidth}
        \vspace{0pt}
        \centering
        \includegraphics[width=0.95\linewidth]
        {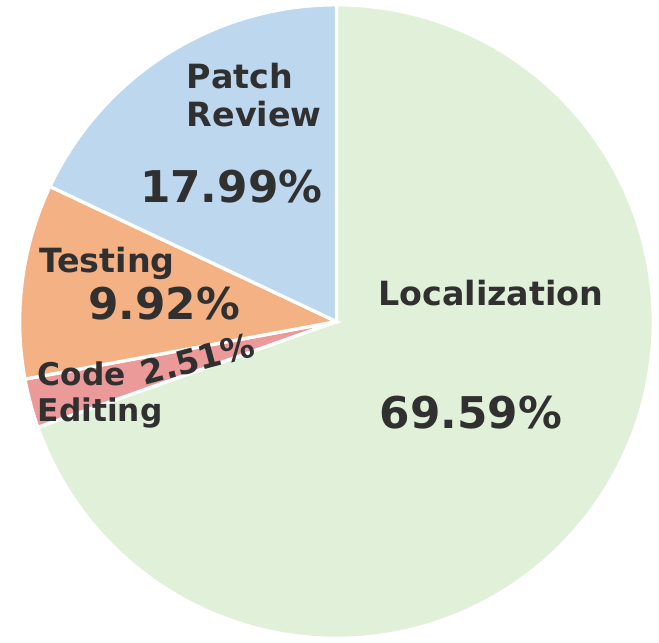}
        \captionof{figure}{Tool ratio.}
        \label{fig:tool_distribution}
    \end{minipage}
    \hfill
    \begin{minipage}[t]{0.75\textwidth}
        \vspace{0pt}
        \centering
        \captionof{table}{Phase-level redundancy analysis on 100 sampled SWE-bench Verified.}
        \label{tab:phase_redundancy}
        \setlength{\tabcolsep}{3pt}

        \resizebox{\linewidth}{!}{
        \begin{tabular}{lccc}
            \toprule
            \textbf{Condition}
            & \textbf{Resolved $\uparrow$}
            & \textbf{Input Tokens $\downarrow$}
            & \textbf{Output Tokens $\downarrow$} \\
            \midrule

            Full Trace
            & 77/100
            & 251,224 {\scriptsize}
            & 6,545 {\scriptsize} \\

            Localized File Provided
            & 78/100
            & 174,844
              {\scriptsize\textcolor{green!50!black}
              {($\downarrow$30.40\%)}}
            & \textbf{5,651}
              {\scriptsize\textcolor{green!50!black}
              {($\downarrow$13.66\%)}} \\

            \bottomrule
        \end{tabular}
        }
    \end{minipage}
\end{figure*}

\textbf{Phase-Level Redundancy.}
Textual overlap captures duplicated content but does not account for observations that become obsolete as the agent transitions between problem-solving phases. As shown in Fig.~\ref{fig:tool_distribution}, localization accounts for 69.59\% of all tool calls, substantially more than code editing (2.51\%), testing (9.92\%), and patch review (17.99\%). A large portion of the interaction trace is therefore generated while the agent searches the repository, inspects candidate files, and diagnoses the reported failure. Although these observations are essential for producing a code modification, their detailed contents may provide limited value once the edit has been applied and the agent proceeds to validation and refinement.


To examine this form of redundancy, we conduct an oracle-style experiment on 100 sampled SWE-bench Verified tasks and compare two settings. In the \emph{Full Trace} setting, the agent follows the standard workflow and retains all tool calls and observations generated during repository localization, code inspection, editing, and validation. In the \emph{Localized File Provided} setting, we directly provide the code location identified during localization and omit the preceding localization trajectory. The agent then starts from the localized code and performs the remaining inspection, editing, and validation steps. Both settings use the same issue description, backbone model, execution environment, and interaction budget. This comparison examines whether the localization result can replace the verbose interaction history used to derive it.

As shown in Table~\ref{tab:phase_redundancy}, directly providing the localized code context achieves a resolution rate comparable to that of the full-trace setting, while reducing input and output token consumption by 30.40\% and 13.66\%, respectively. This result suggests that once the outcome of localization is available, retaining the detailed observations used to derive it provides limited additional value for subsequent repair. It therefore motivates clearing pre-edit observation payloads when the agent transitions from localization to code editing.



These results reveal two forms of redundancy in agent trajectories: content overlap across observations and phase-expired information after code editing. This motivates \scheme \ to remove overlapping content and discard pre-edit observation payloads.

\begin{tcolorbox}[
    colback=gray!20,
    colframe=gray!20,
    width=\columnwidth,
    left=0.05in,
    right=0.05in,
    top=0.05in,
    bottom=0.05in
]
\textbf{Observation 2:}
Agent trajectories contain both cross-step overlap and phase-expired
observations, offering substantial opportunities for context reduction.
\end{tcolorbox}



\section{Method}

\begin{figure}[h]
\centering
\includegraphics[width = .98\textwidth]{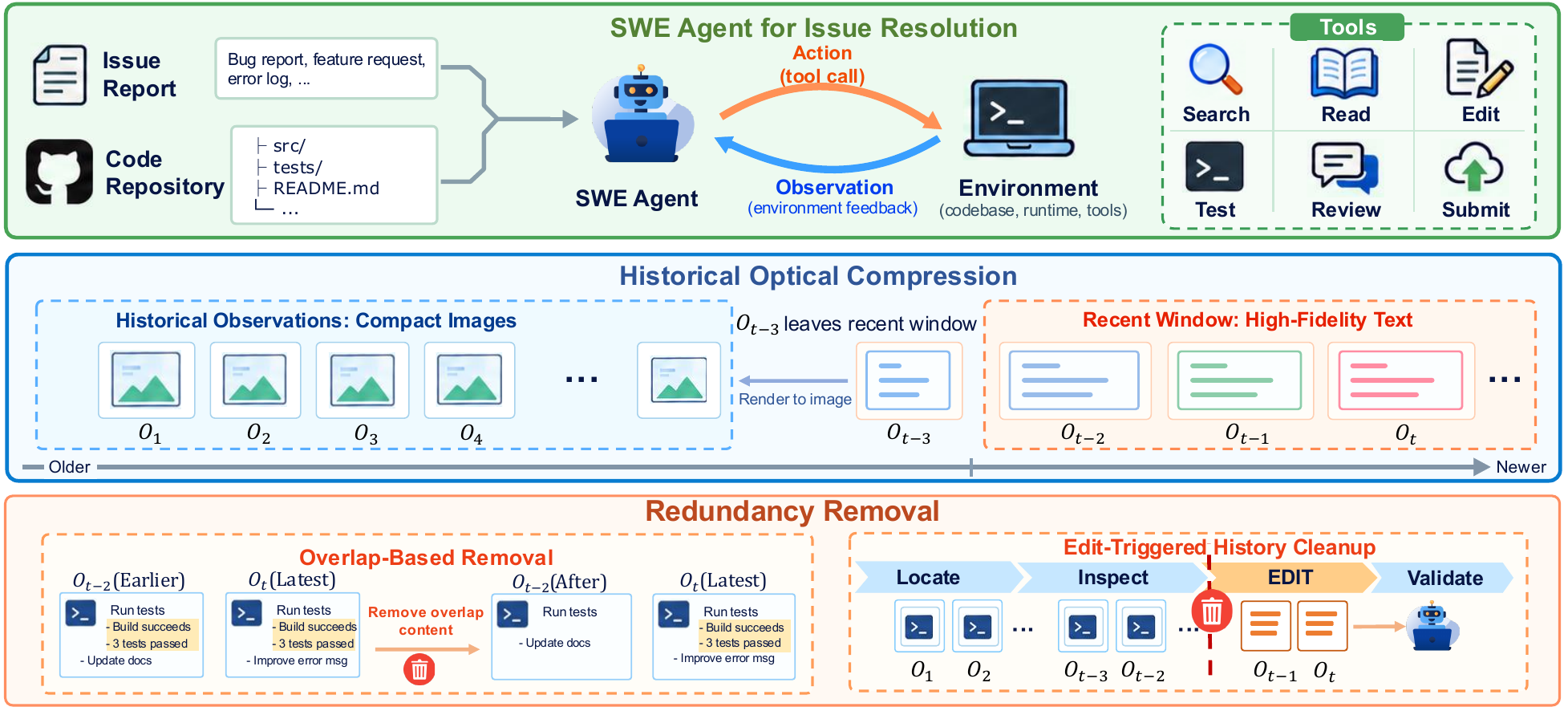}
\caption{Overview of \scheme.}
\label{fig:overview}
\end{figure}

Fig.~\ref{fig:overview} presents the overview of \scheme \ and Algorithm~\ref{alg:tracelite} shows the detailed workflow of \scheme. A SWE agent iteratively interacts with the repository environment through tool calls and observations. \scheme \ manages the resulting interaction trace using two complementary mechanisms. First, \emph{historical optical compression} preserves the most recent observations as high-fidelity text and renders an observation as a compact image once it leaves the recent window. Second, \emph{redundancy removal} eliminates unnecessary information in two ways: overlap-based removal deletes content reproduced by a newer observation while preserving the latest evidence, whereas edit-triggered cleanup removes pre-edit observation payloads when the agent transitions from localization and inspection to editing and validation. Together, these mechanisms reduce the accumulated context while retaining recent evidence, historical tool calls, and repository state needed for subsequent reasoning. The entire process operates online without model training or auxiliary LLM calls.

\subsection{Historical Optical Compression}
\label{sec:historical_optical_compression}

The preliminary study shows that optical rendering can preserve most character-level and semantic information under moderate compression. This finding suggests that observations can be represented as images to reduce the context cost without training a dedicated compressor or invoking an auxiliary LLM. However, optical rendering remains an approximate representation: reducing the image resolution may obscure fine-grained details such as file paths, identifiers, line numbers, version strings, and error messages. Such errors are particularly problematic for recent observations, which usually provide the immediate evidence required to determine the agent's next action.
\begin{table}[h]
    \centering
    \caption{Comparison of full-text and all-image representations on 50 sampled SWE-bench Verified tasks. Values are averages per task, and token changes are calculated relative to the text setting.}
    \label{tab:all_image_comparison}
    \setlength{\tabcolsep}{3pt}
    \begin{tabular}{lcccc}
        \toprule
        \textbf{Representation}
        & \textbf{Resolved $\uparrow$}
        & \textbf{Steps $\downarrow$}
        & \textbf{Input Tokens $\downarrow$}
        & \textbf{Output Tokens $\downarrow$} \\
        \midrule

        Text
        & \textbf{45/50 (90.00\%)}
        & \textbf{12.56}
        & \textbf{187,515}
        & \textbf{5,470} \\

        Observations as Images ($1\times$)
        & 37/50 (74.00\%)
        & 17.86
        & 309,458
          {\scriptsize\textcolor{red}{($\uparrow$65.03\%)}}
        & 7,090
          {\scriptsize\textcolor{red}{($\uparrow$29.62\%)}} \\

        Observations as Images ($4\times$)
        & 29/50 (58.00\%)
        & 20.28
        & 195,415
          {\scriptsize\textcolor{red}{($\uparrow$4.21\%)}}
        & 7,581
          {\scriptsize\textcolor{red}{($\uparrow$38.59\%)}} \\

        \bottomrule
    \end{tabular}
\end{table}

\textbf{Limitations of Rendering All Observations as Images.} To determine whether the entire interaction trajectory can be represented visually, we compare the full-text setting with variants that render all observations as images at compression ratios of $1\times$ and $4\times$. As shown in Table~\ref{tab:all_image_comparison}, the full-text setting resolves 45 of 50 tasks (90.00\%) with an average of 12.56 interaction steps. At $1\times$, the success rate decreases to 74.00\%, while the average number of steps increases to 17.86. This setting also consumes 65.03\% more input tokens because the limited compression cannot offset the visual-token overhead and the additional exploration caused by information loss. Increasing the compression ratio to $4\times$ reduces this token overhead, but further lowers the success rate to 58.00\% and increases the average trajectory length to 20.28 steps. Its output-token consumption is consequently 38.59\% higher than that of the text setting. \textit{Overall, uniformly rendering all observations as images removes the precise recent evidence required for subsequent decisions, leading to longer trajectories and substantially lower issue-resolution performance.}


\textbf{Historical and recent windows.} Motivated by these observations, \scheme \ applies optical compression
according to observation recency. At interaction step $t$, it divides the trajectory $\{O_1,\ldots,O_t\}$ into a recent window $\mathcal{R}_t=\{O_i\mid t-w<i\leq t\}$ and a historical window $\mathcal{H}_t=\{O_i\mid i\leq t-w\}$, where $w$ denotes the recent-window size. Observations in $\mathcal{R}_t$ retain their high-fidelity textual representations, whereas each observation $O_i\in\mathcal{H}_t$ is rendered as a compact image using $\operatorname{Render}(O_i,k)$, where $k$ is the image compression ratio.



Observations in the recent window remain in their original textual form, preserving the exact strings needed for immediate reasoning and tool invocation. Once an observation leaves this window, \scheme \ renders it as an image and includes the image in subsequent multimodal prompts. The corresponding tool calls remain in the textual trajectory so that the agent can still recover which operations it has performed and in what order. The recent-window size $w$ and image compression ratio $k$ are studied in the parameter study (Section~\ref{sec:parameter_study}).





\subsection{Redundancy Removal}

Our preliminary study identifies two forms of redundancy in SWE-agent
trajectories. First, observations produced at different steps may contain
substantial textual overlap. Second, observations collected during
localization and diagnosis may become obsolete after the agent applies a
code modification. \scheme \ addresses these two forms of redundancy
through overlap-based removal and edit-triggered history cleanup.

\textbf{Overlap-Based Removal.}
When a new observation $O_s$ is produced, \scheme \ compares it with
earlier observations in the recent-text window at the line level.
Specifically, each observation is normalized and represented as the set
of its non-empty text lines, denoted by $L(O)$. If $L(O_i)\cap L(O_s)\neq\emptyset$, \scheme \ removes the duplicated lines from the earlier observation $O_i$ while retaining its tool call and non-overlapping lines. The latest observation is preserved because it typically reflects the most recent environment state. This operation is restricted to the recent window to avoid repeatedly modifying the stable historical prefix.


\textbf{Edit-Triggered History Cleanup.}
When the agent performs a code-editing action at step $e$, \scheme \
treats the edit as a transition from localization and diagnosis to validation and refinement. It removes the payloads of all observations
produced before the edit, since their derived information has largely been incorporated into the modified repository state. The historical tool calls and the observation associated with the current edit remain available. The agent then continues from the updated repository state with a substantially smaller context.


Together, overlap-based removal and edit-triggered cleanup eliminate
duplicated and phase-expired information while preserving the trajectory
structure required for subsequent reasoning.

\begin{algorithm*}[h]
\caption{Pipeline of \scheme{}}
\footnotesize
\label{alg:tracelite}

\begin{algorithmic}[1]

\Require Issue instruction $I$, environment $E$, recent-window size
$w$, image compression ratio $k$
\Ensure Resolution result $r$, modified environment $E$

\State $P \gets \Call{MakeInitialPrompt}{I}$
\Comment{Initialize the task prompt}
\State $\mathcal{T} \gets [\,]$, $s \gets 0$
\Comment{Initialize the interaction trajectory and step}

\While{\textbf{true}}

    \State $s \gets s+1$
    \State $\mathcal{C}_s \gets \Call{Serialize}{P,\mathcal{T}}$
    \Comment{Construct the model context}

    \State $m_s \gets
    \operatorname{LLM}_{\mathrm{agent}}(\mathcal{C}_s)$
    \Comment{Perform one agent step}

    \If{$\Call{IsTaskDone}{m_s}$}
        \State $r \gets \Call{ExtractResult}{m_s}$
        \State \Return $(r,E)$
    \EndIf

    \State $(a_s,O_s,E) \gets \Call{ExecTool}{E,m_s}$
    \Comment{Execute the selected tool}

    \State $\mathcal{T} \gets
    \mathcal{T} \oplus [\langle a_s,O_s\rangle]$
    \Comment{Append the action and observation}

\algstore{traceliteMain}
\end{algorithmic}

\vspace{2pt}
\noindent
\colorbox{opticalgreen}{%
\begin{minipage}{\dimexpr\linewidth-2\fboxsep\relax}
\begin{algorithmic}[1]
\algrestore{traceliteMain}

    \Statex \textit{// Compress the observation leaving the recent window}

    \State $h \gets s-w$

    \If{$h \geq 1 \land
    \Call{IsText}{\mathcal{T}[h].O}$}

        \State $\mathcal{T}[h].O \gets
        \Call{Render}{\mathcal{T}[h].O,k}$
        \Comment{Replace text with its image representation}

    \EndIf

\algstore{traceliteOptical}
\end{algorithmic}
\end{minipage}%
}

\vspace{3pt}
\noindent
\colorbox{overlapyellow}{%
\begin{minipage}{\dimexpr\linewidth-2\fboxsep\relax}
\begin{algorithmic}[1]
\algrestore{traceliteOptical}

    \Statex \textit{// Remove redundant contents}

    \If{$\Call{IsEdit}{a_s}$}

        \ForAll{$i \in \{1,\ldots,s-1\}$}

            \State $\mathcal{T}[i].O \gets
            \Call{ClearObservation}{\mathcal{T}[i].O}$
            \Comment{Clear the pre-edit observation}

        \EndFor

    \Else

        \State $\mathcal{R}_s \gets
        \{\max(1,s-w+1),\ldots,s\}$

        \ForAll{$i \in \mathcal{R}_s \setminus \{s\}$}

            \State $\rho_{i,s} \gets
            \Call{OverlapRatio}{\mathcal{T}[i].O,O_s}$

            \If{$\rho_{i,s} > 0$}

                \State $\mathcal{T}[i].O \gets
                \Call{RemoveOverlap}{\mathcal{T}[i].O,O_s}$
                \Comment{Remove overlap content}

            \EndIf

        \EndFor

    \EndIf

\algstore{traceliteOverlap}
\end{algorithmic}
\end{minipage}%
}

\vspace{2pt}

\begin{algorithmic}[1]
\algrestore{traceliteOverlap}

\EndWhile

\end{algorithmic}

\end{algorithm*}

\section{Experimental Setup}
\label{sec:experiment_setup}


\subsection{Benchmarks}


Our primary benchmark is SWE-bench Verified~\cite{jimenez2024swebench}, which contains 500 human-validated issues from real-world Python repositories. Given an issue description and repository snapshot, the agent must identify the cause, modify the relevant files, and produce a patch that passes the benchmark tests. We further evaluate generalizability on DeepSWE~\cite{huang2026deepswe}, which contains original, long-horizon engineering tasks with hidden reference solutions and verifiers. This complementary setting examines whether \scheme\ generalizes beyond public GitHub issues to more complex engineering workflows.

\subsection{Evaluation Metrics}


We evaluate each method using the following metrics: (1) \textbf{Resolved Rate}, measured by the number and percentage of tasks whose generated patches pass all benchmark tests; (2) \textbf{Steps}, measured by the average number of agent--environment interaction steps per task; (3) \textbf{Input Tokens} and \textbf{Output Tokens}, measured by the average cumulative token consumption per task; (4) \textbf{Cost (\$/task)}, calculated from the input and output tokens consumed by the coding agent and any auxiliary API models. To enable comparisons across settings, we additionally normalize the cost against the corresponding Vanilla full-trace baseline, whose normalized cost is set to 1; and (5) \textbf{Compression Overhead}, evaluated in terms of offline training resources, online processing latency per observation, and auxiliary-model token consumption. Token and cost changes are reported relative to the full-trace baseline.


\textbf{Statistical Analysis.}
Since all methods are evaluated on the same benchmark instances, we conduct paired two-sided tests between \scheme\ and each baseline. We use the exact McNemar test for \textbf{Resolved Rate} and the Wilcoxon signed-rank test for \textbf{Steps}, \textbf{Input Tokens}, \textbf{Output Tokens}, and \textbf{Cost (\$/task)}. Differences with $p<0.05$ are considered statistically significant.

\subsection{Baselines}


\begin{itemize}[leftmargin=*, nosep]
    \item \textbf{Vanilla (Full Trace)} retains every tool call and observation as raw text throughout the task without performing any compression.

    \item \textbf{RAG} indexes historical observations and retrieves the most relevant chunks according to the agent's current information need.

    \item \textbf{LLMLingua2}~\cite{pan2024llmlingua2} formulates prompt compression as token classification and uses a separately trained encoder to remove less informative tokens.

    \item \textbf{LongCodeZip}~\cite{shi2025longcodezip} applies a two-stage code compression strategy that ranks function-level chunks using conditional perplexity and selects relevant code under a token budget.

    \item \textbf{SWE-Pruner}~\cite{wang2026swepruner} employs a trained neural skimmer to perform task-aware, line-level pruning according to the agent's current information need.

    \item \textbf{SeeRepo}~\cite{ma2026llmagentscoderepositories} constructs a repository graph over files, classes, and functions to provide structurally relevant repository context for agent reasoning.


    \item \textbf{AgentDiet}~\cite{xiao2026agentdiet} uses an auxiliary reflection LLM to identify and remove redundant, unhelpful, or expired information from the interaction history.
\end{itemize}

\subsection{Implementation}

We use GPT-5.6-Luna~\cite{openai2026gpt56luna} and
Gemini-3.8-Flash~\cite{google2026gemini38flash} as the backbone models. For SWE-bench Verified, we set the reasoning effort to \texttt{high}, the temperature to 0, and the maximum number of interaction steps to 250. We randomly partition its 500 tasks into two disjoint subsets: 100 tasks are used exclusively for the parameter study in Section~\ref{sec:parameter_study}, while the remaining 400 tasks are used for the main evaluation. Based on the parameter study, we set the recent-text window size to three observations and the image compression ratio to $4\times$ for all subsequent experiments. We use GPT-5.6-Luna for the main evaluation and Gemini-3.8-Flash to examine the generalizability of \scheme \ across model families. For DeepSWE~\cite{huang2026deepswe}, we use GPT-5.6-Luna with
\texttt{xhigh} reasoning effort, as described in Section~\ref{sec:generalizability}. Because DeepSWE is a long-horizon benchmark that may require substantially more interactions, we impose no maximum step limit. All methods are implemented within mini-SWE-agent, a lightweight software engineering agent framework that provides tools for repository search, file inspection and editing, test execution, and shell interaction. For each benchmark, all methods use the same agent prompt, tool set, execution environment, reasoning configuration, and interaction budget.




For the baseline implementations, RAG uses UniXCoder~\cite{guo2022unixcoder} to encode queries and historical observations and retrieves the top eight most relevant chunks. LongCodeZip uses Qwen2.5-Coder-7B-Instruct~\cite{hui2024qwen25coder} to estimate conditional perplexity and rank code chunks for compression. SWE-Pruner employs Qwen3-Reranker-0.6B~\cite{zhang2025qwen3embedding} as its neural skimmer to estimate the task relevance of individual code lines. AgentDiet~\cite{xiao2026agentdiet} uses GPT-5-mini as the auxiliary reflection model for identifying redundant, unhelpful, or expired information in the interaction history. We use the same target compression ratio for every observation across all compression baselines to ensure a controlled comparison.


\subsection{Research Questions}



\begin{itemize}[leftmargin=*, nosep]
    \item \textbf{RQ1 (Effectiveness):} Can \scheme \ reduce the token
    consumption of SWE agents while preserving their issue-resolution
    performance?

    \item \textbf{RQ2 (Efficiency):} How does the runtime and resource
    overhead of \scheme \ compare with existing context compression
    approaches?

    \item \textbf{RQ3 (Ablation):} How does each component of
    \scheme \ contribute to overall effectiveness?

    \item \textbf{RQ4 (Parameter Study):} How do the recent-text window size and image compression ratio affect the effectiveness and token consumption of \scheme?

    \item \textbf{RQ5 (Generalizability):} Does \scheme \ generalize
    across different backbone models and software engineering
    benchmarks?

    \item \textbf{RQ6 (Case Study):} How does \scheme \ compress the trace during issue resolution?
\end{itemize}

\section{Experimental Results}

\subsection{Effectiveness (RQ1)}
\label{sec:effectiveness}

We conduct the main experiment on SWE-bench Verified using GPT-5.6-Luna as the backbone model within the mini-SWE-agent framework. We evaluate whether each method can reduce context consumption while preserving the agent's ability to resolve real-world software issues.

\newcommand{\sigyes}[1]{$#1\,(\mathrm{Yes})$}
\newcommand{\signo}[1]{$#1\,(\mathrm{No})$}

\begin{table*}[ht]
    \centering
    \caption{Performance on 400 SWE-bench Verified tasks.
    Interaction steps and token consumption are reported as averages per task. Cost is normalized by the monetary cost of the Vanilla full-trace baseline, whose normalized cost is set to 1. Changes are calculated relative to the full-trace baseline. The row below each method reports the $p$-value and whether its difference from \scheme \ is statistically significant ($p<0.05$). Bold and underlined values denote the best and second-best results, respectively.}
    \label{tab:baseline_results}
    \setlength{\tabcolsep}{1.5pt}

    \resizebox{0.99\textwidth}{!}{
    \begin{tabular}{lccccc}
        \toprule
        \textbf{Method}
        & \textbf{Resolved $\uparrow$}
        & \textbf{Steps $\downarrow$}
        & \textbf{Input Tokens $\downarrow$}
        & \textbf{Output Tokens $\downarrow$}
        & \textbf{Normalized Cost $\downarrow$} \\
        \midrule

        Vanilla (Full Trace)
        & \underline{312/400 (78.00\%)}
        & \underline{14.65}
        & 289,507 {\scriptsize (baseline)}
        & \underline{7,248} {\scriptsize (baseline)}
        & 1.0000 {\scriptsize (baseline)} \\
        \textit{$p$-value}
        & \signo{1}
        & \sigyes{2.17\mathrm{e}{-14}}
        & \sigyes{4.88\mathrm{e}{-39}}
        & \signo{0.1285}
        & \sigyes{2.36\mathrm{e}{-21}} \\
        \midrule

        RAG
        & 305/400 (76.25\%)
        & 16.17
        & 327,896
          {\scriptsize\textcolor{red}{($\uparrow$13.26\%)}}
        & 7,847
          {\scriptsize\textcolor{red}{($\uparrow$8.26\%)}}
        & 1.2061
          {\scriptsize\textcolor{red}{($\uparrow$20.61\%)}}
        \\
        \textit{$p$-value}
        & \signo{0.2799}
        & \sigyes{0.0120}
        & \sigyes{6.35\mathrm{e}{-41}}
        & \signo{0.1363}
        & \sigyes{3.86\mathrm{e}{-35}} \\
        \midrule

        LLMLingua2
        & 280/400 (70.00\%)
        & 21.18
        & 323,401
          {\scriptsize\textcolor{red}{($\uparrow$11.71\%)}}
        & 8,617
          {\scriptsize\textcolor{red}{($\uparrow$18.89\%)}}
        & 1.1067
          {\scriptsize\textcolor{red}{($\uparrow$10.67\%)}}
        \\
        \textit{$p$-value}
        & \sigyes{5.55\mathrm{e}{-6}}
        & \sigyes{5.62\mathrm{e}{-10}}
        & \sigyes{2.87\mathrm{e}{-17}}
        & \signo{0.4829}
        & \sigyes{2.16\mathrm{e}{-6}} \\
        \midrule

        SWE-Pruner
        & 292/400 (73.00\%)
        & 18.45
        & 245,081
          {\scriptsize\textcolor{green!50!black}{($\downarrow$15.35\%)}}
        & 8,867
          {\scriptsize\textcolor{red}{($\uparrow$22.33\%)}}
        & 0.8078
          {\scriptsize\textcolor{green!50!black}{($\downarrow$19.22\%)}}
        \\
        \textit{$p$-value}
        & \sigyes{0.0031}
        & \signo{0.2000}
        & \sigyes{4.51\mathrm{e}{-4}}
        & \sigyes{0.0304}
        & \sigyes{1.83\mathrm{e}{-4}} \\
        \midrule

        LongCodeZip
        & 258/400 (64.50\%)
        & \textbf{12.87}
        & \textbf{130,044}
          {\scriptsize\textcolor{green!50!black}{($\downarrow$55.08\%)}}
        & \textbf{3,529}
          {\scriptsize\textcolor{green!50!black}{($\downarrow$51.31\%)}}
        & \textbf{0.4333}
          {\scriptsize\textcolor{green!50!black}{($\downarrow$56.67\%)}}
        \\
        \textit{$p$-value}
        & \sigyes{2.38\mathrm{e}{-11}}
        & \sigyes{8.19\mathrm{e}{-33}}
        & \sigyes{3.07\mathrm{e}{-24}}
        & \sigyes{8.90\mathrm{e}{-59}}
        & \sigyes{1.54\mathrm{e}{-47}} \\
        \midrule

        AgentDiet
        & 311/400 (77.75\%)
        & 16.03
        & 196,149
          {\scriptsize\textcolor{green!50!black}{($\downarrow$32.25\%)}}
        & 7,712
          {\scriptsize\textcolor{red}{($\uparrow$6.40\%)}}
        & 3.4604
          {\scriptsize\textcolor{red}{($\uparrow$246.04\%)}}
        \\

        \quad $\hookrightarrow$ Auxiliary GPT-5-mini
        & --
        & 11.23 calls/task
        & 83,786
        & 56,679
        & 2.5824 {\scriptsize (included above)} \\

        \textit{$p$-value}
        & \signo{0.8679}
        & \sigyes{0.0014}
        & \signo{0.9025}
        & \signo{0.3493}
        & \sigyes{5.12\mathrm{e}{-67}} \\
        \midrule

        SeeRepo
        & 287/400 (71.75\%)
        & 17.59
        & 259,073
          {\scriptsize\textcolor{green!50!black}{($\downarrow$10.51\%)}}
        & 9,378
          {\scriptsize\textcolor{red}{($\uparrow$29.38\%)}}
        & 0.9683
          {\scriptsize\textcolor{green!50!black}{($\downarrow$3.17\%)}}
        \\
        \textit{$p$-value}
        & \sigyes{3.06\mathrm{e}{-4}}
        & \signo{0.1390}
        & \sigyes{1.37\mathrm{e}{-14}}
        & \sigyes{0.0298}
        & \sigyes{9.91\mathrm{e}{-7}} \\
        \midrule

        \textbf{\scheme{} (Ours)}
        & \textbf{313/400 (78.25\%)}
        & 17.52
        & \underline{193,827}
          {\scriptsize\textcolor{green!50!black}{($\downarrow$33.05\%)}}
        & 7,621
          {\scriptsize\textcolor{red}{($\uparrow$5.14\%)}}
        & \underline{\textbf{0.7870}}
          {\scriptsize\textcolor{green!50!black}{($\downarrow$21.30\%)}}
        \\

        \bottomrule
    \end{tabular}
    }

\end{table*}

\textbf{Issue-Resolution Performance.}
As shown in Table~\ref{tab:baseline_results}, \scheme \ achieves the numerically highest resolved rate, resolving 313 of 400 tasks
(78.25\%). Its performance is statistically comparable to the
full-trace baseline (78.00\%, $p=1.000$), AgentDiet (77.75\%,
$p=0.8679$), and RAG (76.25\%, $p=0.2799$), indicating that compression does not reduce issue-resolution capability. In contrast, SWE-Pruner, LLMLingua2, LongCodeZip, 
and SeeRepo achieve significantly lower resolved rates than \scheme \ ($p<0.05$). LongCodeZip, for example, reduces input tokens by 55.08\% but lowers the resolved rate to 64.50\% ($p=2.38\times10^{-11}$). These results show that minimizing tokens alone is insufficient, an effective method must preserve the information required for subsequent decisions.

\textbf{Token Consumption.} \scheme \ reduces average input-token consumption from 289,507 to 193,827 per task, corresponding to a statistically significant 33.05\% reduction ($p=4.88\times10^{-39}$). It achieves the second-lowest input consumption while maintaining the highest resolved rate. AgentDiet consumes a comparable number of main-agent input tokens
($p=0.9025$), but additionally invokes an auxiliary model. In contrast, RAG and LLMLingua2 increase input consumption by 13.26\% and 11.71\%, respectively, and both differences from \scheme \ are statistically significant.

\begin{figure}[h]
    \subfigure[Input Token/Sample.]{
    \centering
    \includegraphics[width = .22\textwidth]{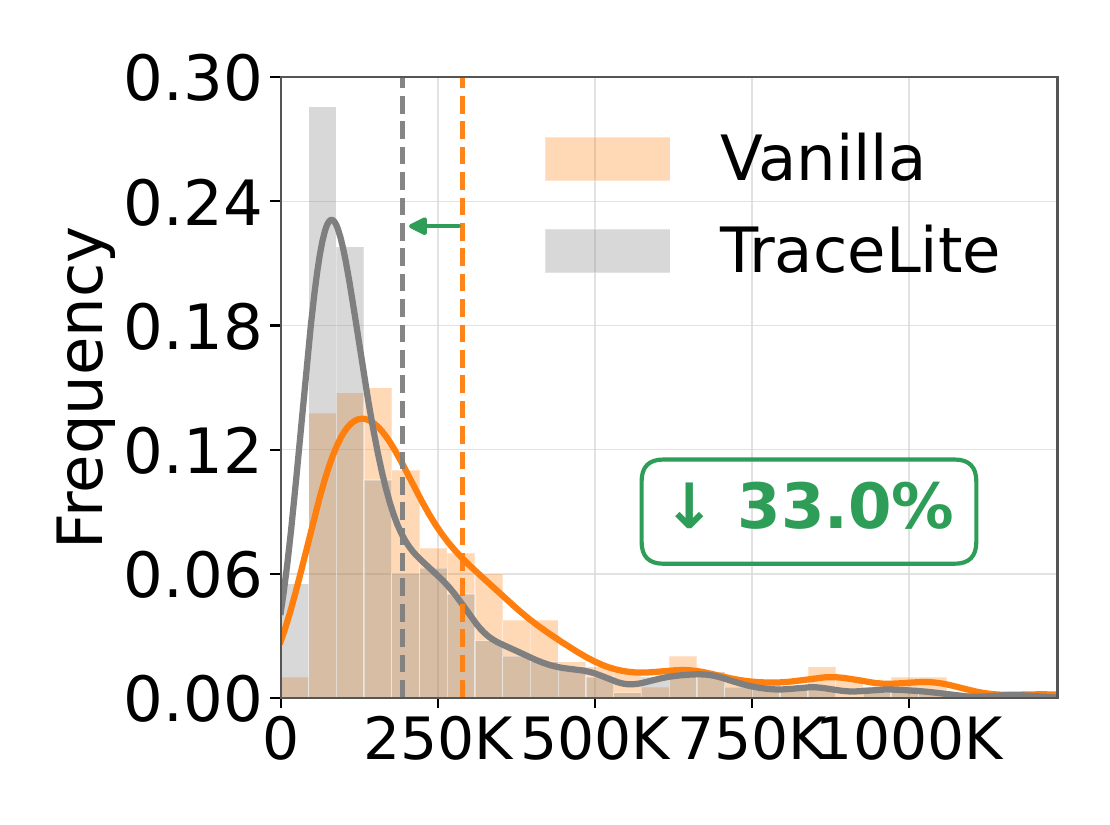}
    }    
    \subfigure[Output Token/Sample.]{
    \centering
    \includegraphics[width = .22\textwidth]{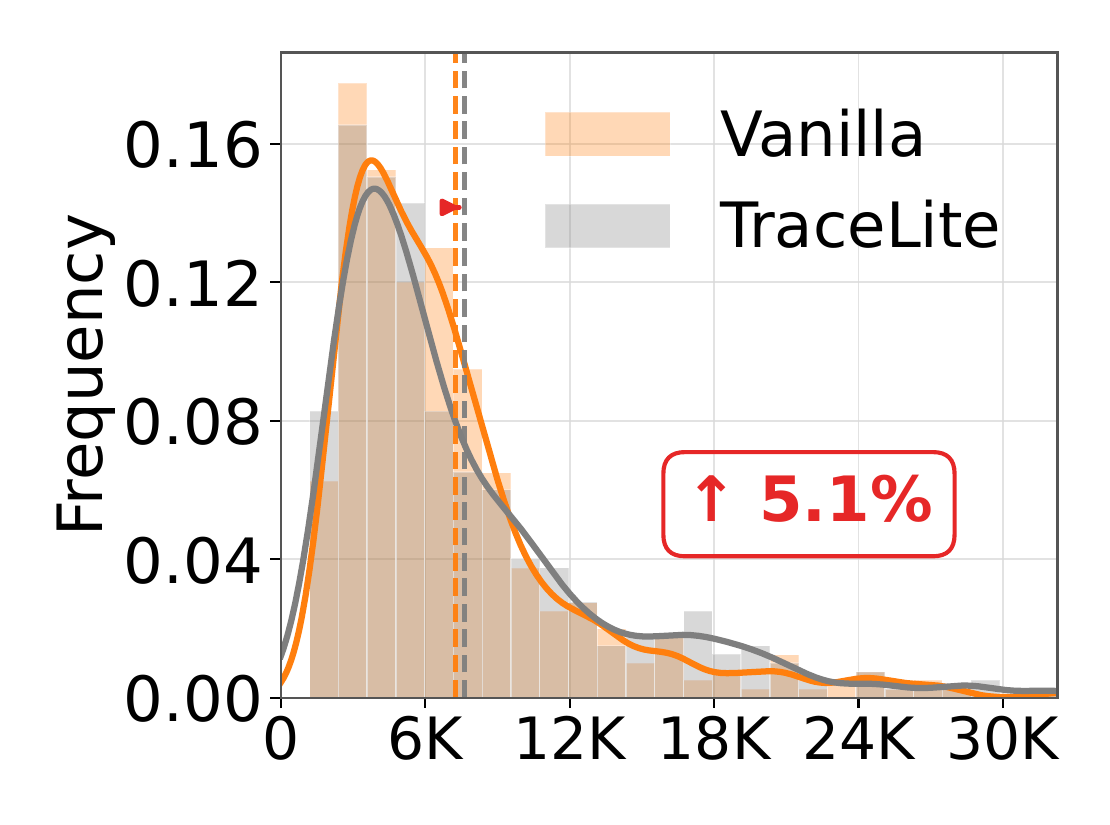}
    }
    \subfigure[Input Token/Step.]{
    \centering
    \includegraphics[width = .22\textwidth]{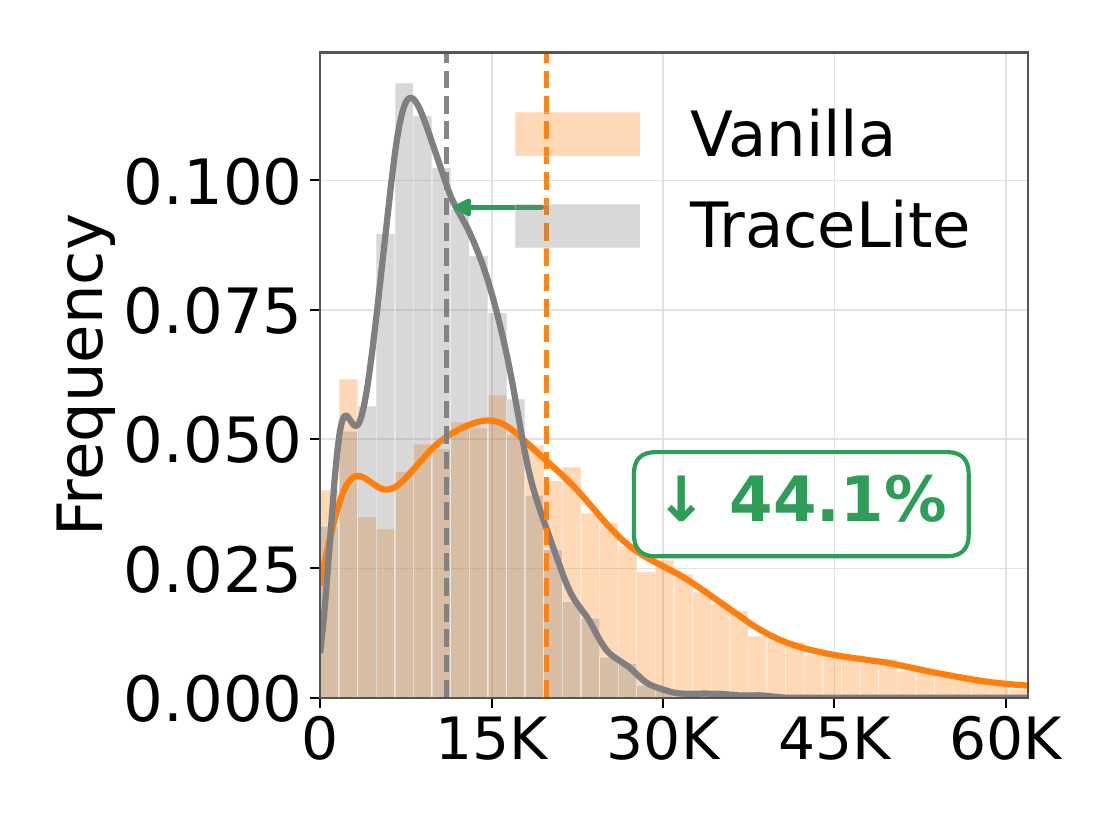}
    }
    \subfigure[Output Token/Step.]{
    \centering
    \includegraphics[width = .22\textwidth]{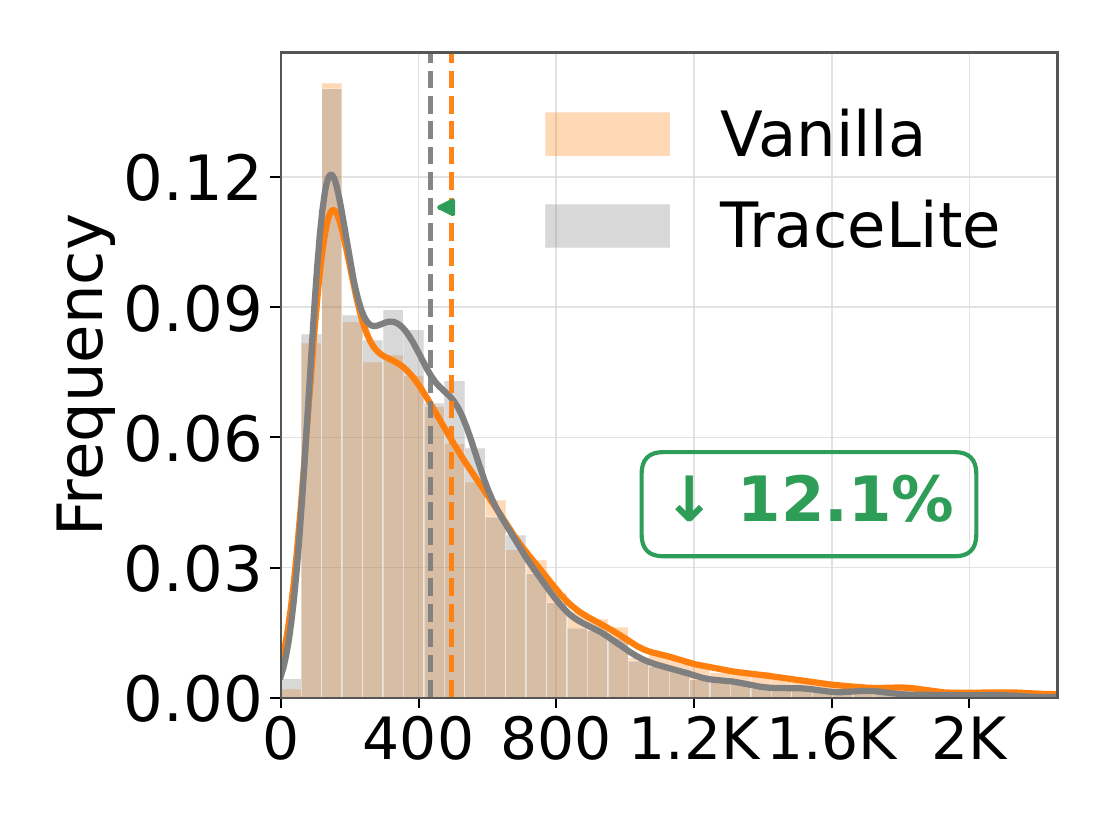}
    }
    \caption{Token distribution of full context and \scheme.}
    \label{fig:token_distribution}
\end{figure}

Figure~\ref{fig:token_distribution} further shows that \scheme \ shifts the input-token distribution toward smaller values and reduces average input consumption per step by approximately 44.1\%. Its per-task reduction is smaller because \scheme \ uses significantly more interaction steps than the full-trace baseline
($p=2.17\times10^{-14}$). Similarly, output consumption per step decreases by approximately 12.1\%, whereas total output tokens increase by only 5.14\%. This increase is not statistically significant ($p=0.1285$), and output tokens remain a small fraction of total token consumption.

\textbf{Cost.} \scheme \ achieves a normalized cost of 0.787, corresponding to a statistically significant reduction of 21.30\% ($p=2.36\times10^{-21}$). LongCodeZip achieves a lower normalized cost of 0.433, but its resolved rate is 13.75 percentage points lower, with a statistically significant performance difference. AgentDiet maintains a comparable resolved rate but reaches a normalized cost of 3.460, representing a 246.04\% increase over the full-trace baseline ($p=5.12\times10^{-67}$), primarily because it invokes GPT-5-mini as an auxiliary reflection model.


Overall, \scheme \ provides the strongest
trade-off among issue-resolution performance, token reduction, and
monetary cost.

\begin{tcolorbox}[
    colback=gray!20,
    colframe=gray!20,
    width=\columnwidth,
    left=0.05in,
    right=0.05in,
    top=0.05in,
    bottom=0.05in
]
\textbf{Answer to RQ1:}
\scheme \ achieves the highest resolved rate while reducing total
input tokens, per-step input tokens, and money cost by 33.05\%,
44.1\%, and 21.30\%, respectively.
\end{tcolorbox}

\subsection{Efficiency (RQ2)}
\label{sec:efficiency}

Beyond reducing context consumption, a compression method may introduce additional costs through model training or online context processing. We therefore evaluate efficiency from both offline and online perspectives. We measure the average processing latency per observation. For offline cost, we report the training time and computational resources required by each method. We additionally report the input and output tokens consumed by separately invoked auxiliary API models. A dash indicates that the corresponding cost is not applicable or that the method does not invoke an auxiliary API model.


\begin{table*}[h]
    \centering
    \caption{Offline training cost and online inference overhead.}
    \label{tab:efficiency}
    {
    \small
    \setlength{\tabcolsep}{1.5pt}
    \begin{tabular}{lccccc}
        \toprule
        & \multicolumn{2}{c}{\textbf{Offline Training Cost}}
        & \multicolumn{3}{c}{\textbf{Online Inference Overhead}} \\
        \cmidrule(lr){2-3}
        \cmidrule(lr){4-6}
        \textbf{Method}
        & \textbf{Time}
        & \textbf{Resources}
        & \shortstack{\textbf{Latency}}
        & \shortstack{\textbf{Input Tokens}}
        & \shortstack{\textbf{Output Tokens}} \\
        \midrule
        RAG
        & --
        & --
        & 0.418
        & --
        & -- \\

        LLMLingua2
        & 23 h
        & 1$\times$ RTX 4090
        & 0.242
        & --
        & -- \\

        LongCodeZip
        & --
        & --
        & 19.438
        & --
        & -- \\

        SWE-Pruner
        & 4 h
        & 8$\times$ A100-80GB
        & 0.707
        & --
        & -- \\

        SeeRepo
        & --
        & --
        & 1.870
        & --
        & -- \\


        AgentDiet
        & --
        & --
        & 4.537
        & 7,461
        & 5047 \\

        \midrule
        \textbf{\scheme}
        & \textbf{--}
        & \textbf{--}
        & 0.438
        & \textbf{--}
        & \textbf{--} \\

        \bottomrule
    \end{tabular}}
\end{table*}

\textbf{Offline Training Cost.}
Table~\ref{tab:efficiency} shows that several methods require substantial offline training. LLMLingua2 requires 23 hours on one RTX 4090, while SWE-Pruner requires four hours on eight A100-80GB GPUs. In contrast, \scheme \ is training-free and can be directly integrated into an existing SWE agent without collecting training data or allocating additional training resources.


\textbf{Online Inference Overhead.}
LLMLingua2 achieves the lowest processing latency at 0.242 seconds per observation, followed by RAG at 0.418 seconds. \scheme \ introduces only 0.438 seconds of additional latency, remaining comparable to these lightweight baselines. It is substantially faster than SWE-Pruner, SeeRepo, AgentDiet, and LongCodeZip. In particular, \scheme \ is approximately $10.4\times$ faster than AgentDiet and $44.4\times$ faster than LongCodeZip.


Most importantly, \scheme \ does not invoke an auxiliary API model during compression and therefore incurs no additional API-token cost. AgentDiet, in contrast, consumes an average of 7,461 input tokens and 5047 output tokens per observation through its reflection model. These additional calls introduce both higher latency and extra token expenditure. Overall, \scheme \ achieves low online processing overhead while avoiding both offline training and auxiliary API inference.

\begin{tcolorbox}[
    colback=gray!20,
    colframe=gray!20,
    width=\columnwidth,
    left=0.05in,
    right=0.05in,
    top=0.05in,
    bottom=0.05in
]
\textbf{Answer to RQ2:}
\scheme \ requires neither offline training nor auxiliary API calls and introduces only 0.438 seconds of latency per observation, providing low offline and online overhead.
\end{tcolorbox}



\subsection{Ablation Study (RQ3)}
\label{sec:ablation}

We conduct an ablation study on 100 tasks randomly sampled from the 400-task evaluation split of SWE-bench Verified. These tasks are disjoint from the 100-task development split used for the parameter study in Section~\ref{sec:parameter_study}. The ablation study examines the contributions of the three components of \scheme: historical optical compression, overlap-based removal, and edit-triggered history cleanup.  We evaluate all eight combinations of the three components, including each component individually, their pairwise combinations, and the complete configuration. Full-history prompting without any compression component serves as the reference configuration.

\begin{table*}[h]
    \centering
    \caption{Ablation study of \scheme\ on 100 sampled SWE-bench
    Verified tasks. ``HOC'' denotes historical optical compression,
    ``OR'' denotes overlap-based removal, and ``EC'' denotes
    edit-triggered cleanup of pre-edit observations. Token changes are
    calculated relative to the full-text configuration $C_0$. Bold and
    underlined values denote the best and second-best results,
    respectively. The $p$-values compare each configuration against the
    complete configuration $C_7$; ``Yes'' denotes statistical
    significance at $p<0.05$.}
    \label{tab:ablation}
    {
    \small
    \setlength{\tabcolsep}{2pt}
  \renewcommand{\arraystretch}{0.82}
    \begin{tabular}{cccccccc}
        \toprule
        \textbf{HOC}
        & \textbf{OR}
        & \textbf{EC}
        &
        & \textbf{Resolved $\uparrow$}
        & \shortstack{\textbf{Steps}$\downarrow$}
        & \shortstack{\textbf{Input Tokens}$\downarrow$}
        & \shortstack{\textbf{Output Tokens}$\downarrow$} \\
        \midrule

        \xmark & \xmark & \xmark
        & $C_0$
        & 77/100 (77.00\%)
        & \textbf{13.93}
        & 251,224 {\scriptsize (baseline)}
        & 6,545 {\scriptsize (baseline)} \\
        & & & \textit{$p$-value}
        & 0.146 (No)
        & 0.001 (Yes)
        & $1.31{\times}10^{-8}$ (Yes)
        & 0.195 (No) \\

        \midrule

        \cmark & \xmark & \xmark
        & $C_1$
        & 79/100 (79.00\%)
        & 15.62
        & 220,465
          {\scriptsize\textcolor{green!50!black}{($\downarrow$12.24\%)}}
        & 6,743
          {\scriptsize\textcolor{red}{($\uparrow$3.03\%)}} \\
        & & & \textit{$p$-value}
        & 0.688 (No)
        & 0.083 (No)
        & $1.42{\times}10^{-6}$ (Yes)
        & 0.592 (No) \\

        \midrule

        \xmark & \cmark & \xmark
        & $C_2$
        & 82/100 (82.00\%)
        & \underline{15.38}
        & 247,295
          {\scriptsize\textcolor{green!50!black}{($\downarrow$1.56\%)}}
        & \textbf{5,143}
          {\scriptsize\textcolor{green!50!black}{($\downarrow$21.42\%)}} \\
        & & & \textit{$p$-value}
        & 1.000 (No)
        & 0.509 (No)
        & $5.38{\times}10^{-10}$ (Yes)
        & 0.051 (No) \\

        \midrule

        \xmark & \xmark & \cmark
        & $C_3$
        & \textbf{84/100 (84.00\%)}
        & 18.77
        & 230,569
          {\scriptsize\textcolor{green!50!black}{($\downarrow$8.22\%)}}
        & 5,988
          {\scriptsize\textcolor{green!50!black}{($\downarrow$8.51\%)}} \\
        & & & \textit{$p$-value}
        & 1.000 (No)
        & 0.004 (Yes)
        & $5.67{\times}10^{-4}$ (Yes)
        & 0.297 (No) \\

        \midrule

        \cmark & \cmark & \xmark
        & $C_4$
        & 78/100 (78.00\%)
        & 15.67
        & 204,153
          {\scriptsize\textcolor{green!50!black}{($\downarrow$18.74\%)}}
        & 6,644
          {\scriptsize\textcolor{red}{($\uparrow$1.51\%)}} \\
        & & & \textit{$p$-value}
        & 0.125 (No)
        & 0.121 (No)
        & 0.001 (Yes)
        & 0.938 (No) \\

        \midrule

        \cmark & \xmark & \cmark
        & $C_5$
        & 80/100 (80.00\%)
        & 17.50
        & \underline{186,040}
          {\scriptsize\textcolor{green!50!black}{($\downarrow$25.95\%)}}
        & 7,207
          {\scriptsize\textcolor{red}{($\uparrow$10.11\%)}} \\
        & & & \textit{$p$-value}
        & 0.508 (No)
        & 0.329 (No)
        & 0.292 (No)
        & 0.108 (No) \\

        \midrule

        \xmark & \cmark & \cmark
        & $C_6$
        & 80/100 (80.00\%)
        & 17.32
        & 223,423
          {\scriptsize\textcolor{green!50!black}{($\downarrow$11.07\%)}}
        & \underline{5,678}
          {\scriptsize\textcolor{green!50!black}{($\downarrow$13.25\%)}} \\
        & & & \textit{$p$-value}
        & 0.344 (No)
        & 0.045 (Yes)
        & $5.18{\times}10^{-6}$ (Yes)
        & 0.009 (Yes) \\

        \midrule

        \cmark & \cmark & \cmark
        & $\mathbf{C_7}$
        & \underline{83/100 (83.00\%)}
        & 17.18
        & \textbf{182,714}
          {\scriptsize\textcolor{green!50!black}{($\downarrow$27.27\%)}}
        & 6,960
          {\scriptsize\textcolor{red}{($\uparrow$6.34\%)}} \\

        \bottomrule
    \end{tabular}}
\end{table*}

Table~\ref{tab:ablation} shows that the complete configuration achieves the best overall trade-off, improving the resolved rate from 77.00\% to 83.00\% while reducing input tokens by 27.27\%. Historical optical compression alone reduces input consumption by 12.24\%, whereas edit-triggered cleanup improves the resolved rate to 84.00\% and reduces input tokens by 8.22\%. Overlap removal alone provides only a modest input-token reduction but yields the largest output-token reduction of 21.42\%. Combining optical compression with overlap removal or edit cleanup reduces input tokens by 18.74\% and 25.95\%, respectively. The resolution-rate differences between the complete method and all ablated configurations are not statistically significant. However, the complete method consumes significantly fewer input tokens than all configurations except \(C_5\). Overall, optical compression primarily reduces representation cost, while overlap removal and edit-triggered cleanup address cross-step and phase-level redundancy, respectively.


\begin{tcolorbox}[
    colback=gray!20,
    colframe=gray!20,
    width=\columnwidth,
    left=0.05in,
    right=0.05in,
    top=0.05in,
    bottom=0.05in
]
\textbf{Answer to RQ3:}
All three components contribute to context reduction. Their combination achieves the largest input-token reduction while preserving issue-resolution performance.
\end{tcolorbox}

\subsection{Parameter Study (RQ4)}
\label{sec:parameter_study}

We investigate two key hyperparameters of \scheme: (1) the number of recent interaction steps retained as text and (2) the compression ratio used to render historical observations as images. These parameters control the trade-off between preserving high-fidelity recent context and reducing the token consumption of historical observations.

\subsubsection{Number of Recent Text Steps.}
We first investigate how many recent observations should be preserved in their original textual form. To reduce the experimental cost, we sample 49 tasks from SWE-bench Verified and vary the recent-text window size from one to five steps. 


\begin{table*}[h]
    \centering
    \caption{Effect of the recent-text window size on 49 sampled tasks
    from SWE-bench Verified. Bold and underlined values denote the
    best and second-best results, respectively.}
    \label{tab:recent_window}
    \setlength{\tabcolsep}{2pt}
    \renewcommand{\arraystretch}{0.82}
    \begin{tabular}{lcccc}
        \toprule
        \textbf{Condition}
        & \textbf{Resolved $\uparrow$}
        & \textbf{Steps $\downarrow$}
        & \textbf{Input Tokens $\downarrow$}
        & \textbf{Output Tokens $\downarrow$} \\
        \midrule

        Vanilla (Full Trace)
        & \underline{41/49 (83.67\%)}
        & \textbf{14.63}
        & 275,464 {\scriptsize (baseline)}
        & 7,576 {\scriptsize (baseline)} \\

        Recent-$1$
        & 40/49 (81.63\%)
        & 20.71
        & 153,634
          {\scriptsize\textcolor{green!50!black}{($\downarrow$44.23\%)}}
        & 7,860
          {\scriptsize\textcolor{red}{($\uparrow$3.75\%)}} \\

        Recent-$2$
        & \underline{41/49 (83.67\%)}
        & 18.35
        & 158,309
          {\scriptsize\textcolor{green!50!black}{($\downarrow$42.53\%)}}
        & 7,635
          {\scriptsize\textcolor{red}{($\uparrow$0.77\%)}} \\

        Recent-$3$
        & \textbf{42/49 (85.71\%)}
        & 17.14
        & \textbf{152,156}
          {\scriptsize\textcolor{green!50!black}{($\downarrow$44.76\%)}}
        & \underline{6,958}
          {\scriptsize\textcolor{green!50!black}{($\downarrow$8.16\%)}} \\

        Recent-$4$
        & 40/49 (81.63\%)
        & \underline{16.51}
        & 165,275
          {\scriptsize\textcolor{green!50!black}{($\downarrow$40.00\%)}}
        & 7,621
          {\scriptsize\textcolor{red}{($\uparrow$0.59\%)}} \\

        Recent-$5$
        & \underline{41/49 (83.67\%)}
        & \textbf{14.63}
        & \underline{153,614}
          {\scriptsize\textcolor{green!50!black}{($\downarrow$44.23\%)}}
        & \textbf{6,891}
          {\scriptsize\textcolor{green!50!black}{($\downarrow$9.04\%)}} \\

        \bottomrule
    \end{tabular}
\end{table*}

As shown in Table~\ref{tab:recent_window}, retaining only the most recent observation substantially reduces input-token consumption, but decreases the pass rate from 83.67\% to 81.63\% and increases the average number of interaction steps from 14.63 to 20.71. Retaining two recent observations recovers the full-history pass rate, although the agent still requires more steps. The three-step window achieves the highest
pass rate of 85.71\% while consuming 152,155 input tokens, representing a 44.76\% reduction relative to the full-history baseline. It also reduces average output-token consumption from 7,576 to 6,958.

Expanding the window beyond three steps does not yield a consistent improvement. The pass rate decreases to 81.63\% with four steps and returns to the full-history result of 83.67\% with five steps. Although the five-step setting produces the fewest output tokens, it does not improve issue-resolution performance. These results suggest that a small recent window is sufficient to preserve the fine-grained information needed for immediate decisions, whereas retaining additional textual observations mainly introduces redundant context. We therefore use three recent text steps.


\subsubsection{Image Compression Ratio.}
We next fix the recent-text window to three steps and investigate the effect of the image compression ratio. Specifically, we evaluate five compression ratios, ranging from $1\times$ to $8\times$, on 100 sampled tasks from SWE-bench Verified. A larger ratio corresponds to a lower image resolution and fewer visual tokens for each historical observation.
\begin{table*}[h]
    \centering
    \caption{Effect of the image compression ratio on 100 sampled tasks from SWE-bench Verified. Bold and underlined values denote the best and second-best results, respectively.}
    \label{tab:compression_ratio}
    \setlength{\tabcolsep}{1pt}
      \renewcommand{\arraystretch}{0.82}
    \begin{tabular}{lcccc}
        \toprule
        \textbf{Method}
        & \textbf{Pass Rate $\uparrow$}
        & \textbf{Steps $\downarrow$}
        & \textbf{Input Tokens $\downarrow$}
        & \textbf{Output Tokens $\downarrow$} \\
        \midrule

        Vanilla (Full Trace)
        & 83/100 (83.00\%)
        & \textbf{13.60}
        & 232,021 {\scriptsize (baseline)}
        & 6,209 {\scriptsize (baseline)} \\

        \scheme{} ($1\times$)
        & 82/100 (82.00\%)
        & 14.52
        & 328,183
          {\scriptsize\textcolor{red}{($\uparrow$41.45\%)}}
        & 6,488
          {\scriptsize\textcolor{red}{($\uparrow$4.50\%)}} \\

        \scheme{} ($2\times$)
        & \underline{84/100 (84.00\%)}
        & 14.21
        & 205,472
          {\scriptsize\textcolor{green!50!black}{($\downarrow$11.44\%)}}
        & \textbf{5,719}
          {\scriptsize\textcolor{green!50!black}{($\downarrow$7.90\%)}} \\


        \scheme{} ($4\times$)
        & \textbf{86/100 (86.00\%)}
        & 14.64
        & 183,719
          {\scriptsize\textcolor{green!50!black}{($\downarrow$20.82\%)}}
        & 6,157
          {\scriptsize\textcolor{green!50!black}{($\downarrow$0.84\%)}} \\


        \scheme{} ($6\times$)
        & 80/100 (80.00\%)
        & \underline{14.08}
        & \textbf{146,383}
          {\scriptsize\textcolor{green!50!black}{($\downarrow$36.91\%)}}
        & \underline{5,885}
          {\scriptsize\textcolor{green!50!black}{($\downarrow$5.22\%)}} \\

        \scheme{} ($8\times$)
        & 77/100 (77.00\%)
        & 14.47
        & \underline{147,521}
          {\scriptsize\textcolor{green!50!black}{($\downarrow$36.42\%)}}
        & 6,076
          {\scriptsize\textcolor{green!50!black}{($\downarrow$2.15\%)}} \\

        \bottomrule
    \end{tabular}
\end{table*}

As shown in Table~\ref{tab:compression_ratio}, the $1\times$ setting increases input token cost by 41.45\%, indicating that the reduction in image resolution is insufficient to offset the resulting visual-token overhead. Increasing the ratio to $2\times$ reduces input tokens by 11.44\% while improving the pass rate from 83\% to 84\%. The $4\times$ setting achieves the highest pass rate of 86\% and reduces input-token cost by 20.82\%. More aggressive compression provides greater token savings but begins to degrade issue-resolution performance. Although $6\times$ achieves the lowest input token cost, its pass rate decreases to 80\%. At $8\times$, the pass rate further drops to 77\%, suggesting that important fine-grained information is lost at excessively low resolutions. Overall, $4\times$ provides the best balance between token reduction and task performance, and we therefore use it as the default compression ratio of \scheme.

\begin{tcolorbox}[
    colback=gray!20,
    colframe=gray!20,
    width=\columnwidth,
    left=0.05in,
    right=0.05in,
    top=0.05in,
    bottom=0.05in
]
\textbf{Answer to RQ4:}
A three step recent-text window and a $4\times$ image compression ratio provide the best balance between issue resolution performance and token efficiency.
\end{tcolorbox}

\subsection{Generalizability (RQ5)}
\label{sec:generalizability}

We evaluate the generalizability of \scheme\ from two perspectives: backbone models and benchmarks. For backbone generalization, we replace GPT-5.6-Luna with Gemini-3.8-Flash~\cite{google2026gemini38flash} and evaluate all methods on 50 tasks sampled from SWE-bench Verified. The reasoning effort is set to \texttt{high}. For benchmark generalization, we further evaluate \scheme\ on DeepSWE~\cite{huang2026deepswe} using GPT-5.6-Luna while retaining the same agent configuration and default compression parameters.

\begin{table*}[h]
    \centering
    \caption{Generalizability across backbone models. All methods are
    evaluated on 50 SWE-bench Verified tasks using Gemini-3.8-Flash.
    Cost is normalized by the monetary cost of the Vanilla full-trace
    baseline, whose normalized cost is set to 1. Bold and underlined
    values denote the best and second-best results, respectively.}
    \label{tab:generalizability_backbone}
    \setlength{\tabcolsep}{2pt}

    \resizebox{0.99\textwidth}{!}{
    \begin{tabular}{lccccc}
        \toprule
        \textbf{Method}
        & \textbf{Resolved $\uparrow$}
        & \textbf{Steps $\downarrow$}
        & \textbf{Input Tokens $\downarrow$}
        & \textbf{Output Tokens $\downarrow$}
        & \textbf{Normalized Cost $\downarrow$} \\
        \midrule

        Vanilla (Full Trace)
        & 47/50 (94.00\%)
        & 125.22
        & 5,373,680 {\scriptsize (baseline)}
        & 68,258 {\scriptsize (baseline)}
        & 1.0000 {\scriptsize (baseline)} \\

        SWE-Pruner
        & \underline{48/50 (96.00\%)}
        & \underline{70.20}
        & \underline{2,923,141}
          {\scriptsize\textcolor{green!50!black}
          {($\downarrow$45.60\%)}}
        & \underline{41,783}
          {\scriptsize\textcolor{green!50!black}
          {($\downarrow$38.79\%)}}
        & 0.8090
          {\scriptsize\textcolor{green!50!black}
          {($\downarrow$19.10\%)}} \\

        LLMLingua2
        & 47/50 (94.00\%)
        & 72.36
        & 3,249,537
          {\scriptsize\textcolor{green!50!black}
          {($\downarrow$39.53\%)}}
        & 45,180
          {\scriptsize\textcolor{green!50!black}
          {($\downarrow$33.81\%)}}
        & \underline{0.6427}
          {\scriptsize\textcolor{green!50!black}
          {($\downarrow$35.73\%)}} \\

        LongCodeZip
        & 47/50 (94.00\%)
        & 74.22
        & 3,431,914
          {\scriptsize\textcolor{green!50!black}
          {($\downarrow$36.13\%)}}
        & 44,203
          {\scriptsize\textcolor{green!50!black}
          {($\downarrow$35.24\%)}}
        & 0.6475
          {\scriptsize\textcolor{green!50!black}
          {($\downarrow$35.25\%)}} \\

        AgentDiet
        & 44/50 (88.00\%)
        & \textbf{61.06}
        & 4,646,082
          {\scriptsize\textcolor{green!50!black}
          {($\downarrow$13.54\%)}}
        & \textbf{38,813}
          {\scriptsize\textcolor{green!50!black}
          {($\downarrow$43.14\%)}}
        & 0.9272
          {\scriptsize\textcolor{green!50!black}
          {($\downarrow$7.28\%)}} \\

        \quad $\hookrightarrow$ Auxiliary GPT-5-mini
        & --
        & 12.00 calls/task
        & 89,605
        & 17,563
        & 0.0440 {\scriptsize (included above)} \\

        \midrule

        \textbf{\scheme{} (Ours)}
        & \textbf{49/50 (98.00\%)}
        & 77.67
        & \textbf{2,558,259}
          {\scriptsize\textcolor{green!50!black}
          {($\downarrow$52.39\%)}}
        & 41,805
          {\scriptsize\textcolor{green!50!black}
          {($\downarrow$38.75\%)}}
        & \textbf{0.6217}
          {\scriptsize\textcolor{green!50!black}
          {($\downarrow$37.83\%)}} \\

        \bottomrule
    \end{tabular}
    }
\end{table*}


\textbf{Generalization across Backbone Models.}
As shown in Table~\ref{tab:generalizability_backbone}, the full-trace baseline with Gemini-3.8-Flash resolves 47 of 50 tasks (94.00\%). \scheme \ achieves the highest resolved rate of 98.00\%, while reducing interaction steps, input tokens, and output tokens by 37.98\%, 52.39\%, and 38.75\%, respectively. It also obtains the lowest normalized cost of 0.6217, corresponding to a 37.83\% reduction over the full-trace baseline. SWE-Pruner achieves the second-highest resolved rate of 96.00\%, but consumes 14.26\% more input tokens and costs 30.12\% more than \scheme. Although AgentDiet reduces the main agent's output-token
consumption, its resolved rate decreases to 88.00\%, and its auxiliary GPT-5-mini calls result in a normalized cost of 0.9272. These results show that \scheme \ generalizes effectively to a different backbone model while providing the best overall trade-off between resolution performance and inference cost.



\textbf{Generalization across Benchmarks.}
We further evaluate \scheme \ on DeepSWE~\cite{huang2026deepswe} using GPT-5.6-Luna with \texttt{xhigh} reasoning effort and no maximum interaction-step limit. We retain the same agent framework and default compression parameters to examine whether \scheme \ generalizes beyond SWE-bench Verified. As shown in
Table~\ref{tab:generalizability_benchmark}, \scheme \ preserves the 60.00\% resolved rate of the full-trace baseline while reducing input tokens, output tokens, and normalized cost by 50.99\%, 6.77\%, and 44.31\%, respectively. Despite requiring slightly more interaction steps (94.79 versus 89.34), it achieves the lowest input-token consumption and cost among all methods. LongCodeZip obtains the second-lowest cost but resolves one fewer task and increases output-token consumption by 27.06\%. AgentDiet achieves the highest resolved rate of 62.00\%, but its auxiliary GPT-5-mini calls increase its normalized cost to 1.6403. SWE-Pruner and LLMLingua2 consume more tokens and incur higher costs while achieving lower resolved rates. Overall, \scheme \ provides the strongest efficiency gains on DeepSWE without reducing the full-trace baseline's issue-resolution performance.

\begin{table*}[h]
    \centering
    \caption{Generalizability across software engineering benchmarks.
    All methods are evaluated on 50 DeepSWE tasks using GPT-5.6-Luna.
    Cost is normalized by the monetary cost of the Vanilla full-trace
    baseline, whose normalized cost is set to 1. Bold and underlined
    values denote the best and second-best results, respectively.}
    \label{tab:generalizability_benchmark}
    \setlength{\tabcolsep}{2pt}

    \resizebox{0.99\textwidth}{!}{
    \begin{tabular}{lccccc}
        \toprule
        \textbf{Method}
        & \textbf{Resolved $\uparrow$}
        & \textbf{Steps $\downarrow$}
        & \textbf{Input Tokens $\downarrow$}
        & \textbf{Output Tokens $\downarrow$}
        & \textbf{Normalized Cost $\downarrow$} \\
        \midrule

        Vanilla (Full Trace)
        & \underline{30/50 (60.00\%)}
        & \textbf{89.34}
        & 11,491,143 {\scriptsize (baseline)}
        & 156,832 {\scriptsize (baseline)}
        & 1.0000 {\scriptsize (baseline)} \\

        SWE-Pruner
        & 27/50 (54.00\%)
        & 112.88
        & 13,748,860
          {\scriptsize\textcolor{red}{($\uparrow$19.65\%)}}
        & 182,297
          {\scriptsize\textcolor{red}{($\uparrow$16.24\%)}}
        & 1.1597
          {\scriptsize\textcolor{red}{($\uparrow$15.97\%)}} \\

        LLMLingua2
        & 19/50 (38.00\%)
        & 294.54
        & 31,259,996
          {\scriptsize\textcolor{red}{($\uparrow$172.04\%)}}
        & 251,366
          {\scriptsize\textcolor{red}{($\uparrow$60.28\%)}}
        & 1.5579
          {\scriptsize\textcolor{red}{($\uparrow$55.79\%)}} \\

        LongCodeZip
        & 29/50 (58.00\%)
        & 99.00
        & 10,295,587
          {\scriptsize\textcolor{green!50!black}{($\downarrow$10.40\%)}}
        & 199,275
          {\scriptsize\textcolor{red}{($\uparrow$27.06\%)}}
        & \underline{0.6550}
          {\scriptsize\textcolor{green!50!black}{($\downarrow$34.50\%)}} \\

        AgentDiet
        & \textbf{31/50 (62.00\%)}
        & 97.34
        & \underline{8,324,333}
          {\scriptsize\textcolor{green!50!black}{($\downarrow$27.56\%)}}
        & \textbf{146,052}
          {\scriptsize\textcolor{green!50!black}{($\downarrow$6.87\%)}}
        & 1.6403
          {\scriptsize\textcolor{red}{($\uparrow$64.03\%)}} \\

        \quad $\hookrightarrow$ Auxiliary GPT-5-mini
        & --
        & 69.46 calls/task
        & 561,420
        & 401,106
        & 0.7563 {\scriptsize (included above)} \\

        \midrule

        \textbf{\scheme{} (Ours)}
        & \underline{30/50 (60.00\%)}
        & \underline{94.79}
        & \textbf{5,631,724}
          {\scriptsize\textcolor{green!50!black}{($\downarrow$50.99\%)}}
        & \underline{146,209}
          {\scriptsize\textcolor{green!50!black}{($\downarrow$6.77\%)}}
        & \textbf{0.5569}
          {\scriptsize\textcolor{green!50!black}{($\downarrow$44.31\%)}} \\

        \bottomrule
    \end{tabular}
    }
\end{table*}



\begin{tcolorbox}[
    colback=gray!20,
    colframe=gray!20,
    width=\columnwidth,
    left=0.05in,
    right=0.05in,
    top=0.05in,
    bottom=0.05in
]
\textbf{Answer to RQ5:}
\scheme\ consistently preserves effectiveness while reducing context consumption across different backbone models and benchmarks.
\end{tcolorbox}

\subsection{Case Study (RQ6)}
\begin{figure}[ht]
\centering
\includegraphics[width = .99\textwidth]{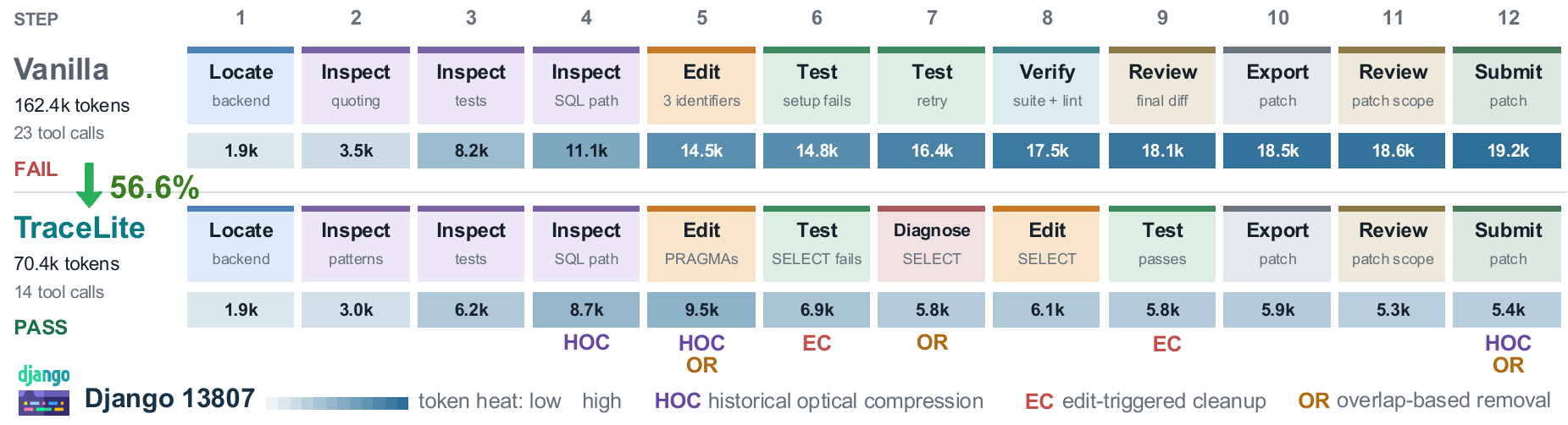}
\caption{Case study on Django issue 13807.}
\label{fig:example}
\end{figure}

As shown in Fig.~\ref{fig:example}, both agents begin by locating the relevant backend and inspecting the quoting logic, tests, and SQL path for Django issue 13807. The Vanilla agent edits three identifiers but encounters a test-setup failure, leading to repeated testing, verification, and patch review; it ultimately fails after 23 tool calls and 162.4K input tokens. In contrast, \scheme \ controls context growth through historical optical compression (\textsc{HOC}), overlap-based removal (\textsc{OR}), and edit-triggered cleanup (\textsc{EC}). After the initial test failure, the compact trajectory retains sufficient recent evidence for the agent to diagnose the incorrect \texttt{SELECT} handling, revise the patch, and pass the subsequent test. \scheme \ successfully resolves the issue using 14 tool calls and 70.4K input tokens, reducing token consumption by 56.6\%. This case illustrates that removing redundant and phase-expired context can improve efficiency while preserving the evidence needed for iterative debugging.

\begin{tcolorbox}[
    colback=gray!20,
    colframe=gray!20,
    width=\columnwidth,
    left=0.05in,
    right=0.05in,
    top=0.05in,
    bottom=0.05in
]
\textbf{Answer to RQ6:}
\scheme \ effectively reduce context growth during issue resolution through optical compression and redundancy removal.
\end{tcolorbox}

\section{Threats to Validity}

(1) \textit{Generalization Across Agent Frameworks.}
Due to the computational and API costs of evaluating complete trajectories, we conduct experiments only with mini-SWE-agent, which may limit generalization to other agent frameworks. Nevertheless, mini-SWE-agent is a competitive framework whose iterative workflow of repository search, file inspection, editing, and testing is representative of modern coding agents. Moreover, \scheme \ processes tool calls and observations without relying on framework-specific prompts or architectures. Evaluations across multiple backbone models and benchmarks further mitigate this threat. (2) \textit{Data Leakage and Benchmark Contamination.} Because SWE-bench Verified contains public GitHub issues and patches, its solutions may occur in model pretraining data and inflate absolute resolved rates. We mitigate this threat by additionally evaluating on DeepSWE~\cite{huang2026deepswe}, whose tasks are written from scratch and whose solutions and verifier tests remain hidden from the agent. For SWE-bench Verified, we follow the official protocol without exposing gold patches or evaluation results. Since all methods use identical models and instances, contamination is unlikely to systematically favor \scheme, although its influence cannot be completely eliminated. (3) \textit{Correctness of Patches.} Both benchmarks assess patches using executable tests or program-based verifiers, which cannot guarantee correctness for every possible input. DeepSWE reduces this risk using hand-written functional verifiers that evaluate observable behavior and accept alternative implementations; its independent audit reports only 1.4\% disagreement between the verifier and an LLM judge~\cite{huang2026deepswe}. Because all methods are evaluated with the same verifiers, potential imperfections may affect absolute pass rates but are unlikely to alter their relative comparison.

\section{Related Work}

\textbf{LLM-Based Issue Resolution}.
LLM-based agents have been widely applied to repository-level issue resolution. SWE-agent~\cite{yang2024sweagent} enables models to iteratively inspect repositories, modify code, and validate patches, while AutoCodeRover~\cite{zhang2024autocoderover} incorporates structure-aware code search and Agentless~\cite{xia2024agentless} decomposes resolution into localization, repair, and validation. Subsequent work improves issue localization through graph-guided reasoning~\cite{liu2026GraphLocator}, tool-interactive training~\cite{ma2026enhancing}, and test-driven execution~\cite{liu2026issueexec}, or enhances resolution through ensemble reasoning~\cite{zhao2026ensemble} and turn-control strategies~\cite{gao2026more}. Meng et al.~\cite{meng2026bugfix} further examine the capabilities and limitations of current bug-fixing agents. Unlike these studies, \scheme \ focuses on compressing growing interaction traces and is complementary to advances in agent reasoning and issue resolution.

\textbf{Token Compression for LLM Agents}.
Existing approaches include general-purpose and coding-specific methods. LLMLingua2~\cite{pan2024llmlingua2} performs general prompt compression using a trained token classifier. For coding agents, AgentDiet~\cite{xiao2026agentdiet} invokes a reflection LLM to remove redundant or expired trajectory content, while LongCodeZip~\cite{shi2025longcodezip} uses conditional perplexity for function- and block-level code pruning. SWE-Pruner~\cite{wang2026swepruner} employs a trained neural skimmer for line-level pruning. These code-oriented methods do not directly handle heterogeneous tool observations. SeeRepo~\cite{ma2026llmagentscoderepositories} improves navigation through visual repository graphs but augments rather than compresses interaction histories. \scheme \ combines optical compression with overlap-based removal and edit-triggered cleanup, requiring neither model training nor auxiliary model calls.



\textbf{Coding Agents and Context Management}.
Coding agents solve software engineering tasks through iterative reasoning, tool use, and environment feedback. SWE-agent~\cite{yang2024sweagent} manages context using windowed file views and observation removal, whereas mini-SWE-agent~\cite{yang2024sweagent} maintains a simple append-only interaction history. OpenHands~\cite{wang2024openhands} provides context condensers that summarize earlier interactions. Commercial and open-source agents adopt similar mechanisms: Claude Code~\cite{anthropic2026claudecode}, Codex~\cite{openai2026codex}, TRAE~\cite{gao2025trae}, and OpenCode~\cite{opencode2026compaction} compact or summarize historical context during long-running tasks. These approaches may discard fine-grained information or require additional model inference. In contrast, \scheme \ retains recent observations as text and compresses historical context through optical re-encoding and deterministic redundancy removal.


\section{Conclusion}

We present \scheme, a training-free framework that compresses SWE-agent traces through recent-text preservation, historical optical compression, and redundancy removal. Evaluations across two benchmarks and backbone models demonstrate its generalizability. On SWE-bench Verified, \scheme \ achieves a 78.25\% resolved rate while reducing input tokens and API cost by 33.05\% and 21.30\%, respectively, with only 0.438 seconds of latency per observation. These results demonstrate that \scheme \ improves SWE-agent efficiency without offline training, auxiliary LLM calls, or degraded issue-resolution capability.

\bibliographystyle{ACM-Reference-Format}
\bibliography{sample-base}


\begin{thebibliography}{28}


\ifx \showCODEN    \undefined \def \showCODEN     #1{\unskip}     \fi
\ifx \showISBNx    \undefined \def \showISBNx     #1{\unskip}     \fi
\ifx \showISBNxiii \undefined \def \showISBNxiii  #1{\unskip}     \fi
\ifx \showISSN     \undefined \def \showISSN      #1{\unskip}     \fi
\ifx \showLCCN     \undefined \def \showLCCN      #1{\unskip}     \fi
\ifx \shownote     \undefined \def \shownote      #1{#1}          \fi
\ifx \showarticletitle \undefined \def \showarticletitle #1{#1}   \fi
\ifx \showURL      \undefined \def \showURL       {\relax}        \fi
\providecommand\bibfield[2]{#2}
\providecommand\bibinfo[2]{#2}
\providecommand\natexlab[1]{#1}
\providecommand\showeprint[2][]{arXiv:#2}

\bibitem[{Anthropic}(2026)]%
        {anthropic2026claudecode}
\bibfield{author}{\bibinfo{person}{{Anthropic}}.} \bibinfo{year}{2026}\natexlab{}.
\newblock \bibinfo{title}{Claude Code: Interactive Mode and Commands}.
\newblock \bibinfo{howpublished}{Claude Code Documentation}.
\newblock
\shownote{Accessed: 2026-09-23}.
\newblock
\urldef\tempurl%
\url{https://code.claude.com/docs/en/commands}
\showURL{%
\tempurl}


\bibitem[Gao and Peng(2026)]%
        {gao2026more}
\bibfield{author}{\bibinfo{person}{Pengfei Gao} {and} \bibinfo{person}{Chao Peng}.} \bibinfo{year}{2026}\natexlab{}.
\newblock \showarticletitle{More with Less: An Empirical Study of Turn-Control Strategies for Efficient Coding Agents}. In \bibinfo{booktitle}{\emph{Proceedings of the 2026 IEEE/ACM 48th International Conference on Software Engineering}} \emph{(\bibinfo{series}{ICSE '26})}. \bibinfo{publisher}{Association for Computing Machinery}, \bibinfo{address}{New York, NY, USA}, \bibinfo{pages}{956–967}.
\newblock
\showISBNx{9798400720253}
\href{https://doi.org/10.1145/3744916.3773250}{doi:\nolinkurl{10.1145/3744916.3773250}}


\bibitem[Gao et~al\mbox{.}(2025)]%
        {gao2025trae}
\bibfield{author}{\bibinfo{person}{Pengfei Gao}, \bibinfo{person}{Zhao Tian}, \bibinfo{person}{Xiangxin Meng}, \bibinfo{person}{Xinchen Wang}, \bibinfo{person}{Ruida Hu}, \bibinfo{person}{Yuanan Xiao}, \bibinfo{person}{Yizhou Liu}, \bibinfo{person}{Zhao Zhang}, \bibinfo{person}{Junjie Chen}, \bibinfo{person}{Cuiyun Gao}, {et~al\mbox{.}}} \bibinfo{year}{2025}\natexlab{}.
\newblock \showarticletitle{Trae agent: An llm-based agent for software engineering with test-time scaling}.
\newblock \bibinfo{journal}{\emph{arXiv preprint arXiv:2507.23370}} (\bibinfo{year}{2025}).
\newblock


\bibitem[{Google DeepMind}(2026)]%
        {google2026gemini38flash}
\bibfield{author}{\bibinfo{person}{{Google DeepMind}}.} \bibinfo{year}{2026}\natexlab{}.
\newblock \bibinfo{title}{{Gemini 3.8 Flash} Model Card}.
\newblock \bibinfo{howpublished}{\url{https://deepmind.google/models/model-cards/gemini-3-8-flash/}}.
\newblock
\shownote{Accessed: 2026-9-16}.
\newblock


\bibitem[Guo et~al\mbox{.}(2022)]%
        {guo2022unixcoder}
\bibfield{author}{\bibinfo{person}{Daya Guo}, \bibinfo{person}{Shuai Lu}, \bibinfo{person}{Nan Duan}, \bibinfo{person}{Yanlin Wang}, \bibinfo{person}{Ming Zhou}, {and} \bibinfo{person}{Jian Yin}.} \bibinfo{year}{2022}\natexlab{}.
\newblock \showarticletitle{Unixcoder: Unified cross-modal pre-training for code representation}. In \bibinfo{booktitle}{\emph{Proceedings of the 60th Annual Meeting of the Association for Computational Linguistics (Volume 1: Long Papers)}}. \bibinfo{pages}{7212--7225}.
\newblock


\bibitem[Huang et~al\mbox{.}(2026)]%
        {huang2026deepswe}
\bibfield{author}{\bibinfo{person}{Wenqi Huang}, \bibinfo{person}{Charley Lee}, \bibinfo{person}{Leonard Tng}, {and} \bibinfo{person}{Serena Ge}.} \bibinfo{year}{2026}\natexlab{}.
\newblock \bibinfo{title}{DeepSWE: Measuring Frontier Coding Agents on Original, Long-Horizon Engineering Tasks}.
\newblock
\showeprint[arxiv]{2607.07946}~[cs.SE]
\urldef\tempurl%
\url{https://arxiv.org/abs/2607.07946}
\showURL{%
\tempurl}


\bibitem[Hui et~al\mbox{.}(2024)]%
        {hui2024qwen25coder}
\bibfield{author}{\bibinfo{person}{Binyuan Hui}, \bibinfo{person}{Jian Yang}, \bibinfo{person}{Zeyu Cui}, \bibinfo{person}{Jiaxi Yang}, \bibinfo{person}{Dayiheng Liu}, \bibinfo{person}{Lei Zhang}, \bibinfo{person}{Tianyu Liu}, \bibinfo{person}{Jiajun Zhang}, \bibinfo{person}{Bowen Yu}, \bibinfo{person}{Keming Lu}, {et~al\mbox{.}}} \bibinfo{year}{2024}\natexlab{}.
\newblock \showarticletitle{Qwen2. 5-coder technical report}.
\newblock \bibinfo{journal}{\emph{arXiv preprint arXiv:2409.12186}} (\bibinfo{year}{2024}).
\newblock


\bibitem[Jimenez et~al\mbox{.}(2024)]%
        {jimenez2024swebench}
\bibfield{author}{\bibinfo{person}{Carlos~E Jimenez}, \bibinfo{person}{John Yang}, \bibinfo{person}{Alexander Wettig}, \bibinfo{person}{Shunyu Yao}, \bibinfo{person}{Kexin Pei}, \bibinfo{person}{Ofir Press}, {and} \bibinfo{person}{Karthik~R Narasimhan}.} \bibinfo{year}{2024}\natexlab{}.
\newblock \showarticletitle{{SWE}-bench: Can Language Models Resolve Real-world Github Issues?}. In \bibinfo{booktitle}{\emph{The Twelfth International Conference on Learning Representations}}.
\newblock
\urldef\tempurl%
\url{https://openreview.net/forum?id=VTF8yNQM66}
\showURL{%
\tempurl}


\bibitem[Liu et~al\mbox{.}(2026a)]%
        {liu2026issueexec}
\bibfield{author}{\bibinfo{person}{Jiawei Liu}, \bibinfo{person}{Yun Lin}, \bibinfo{person}{Chenyan Liu}, \bibinfo{person}{Yu Qian}, \bibinfo{person}{Yiming Liu}, \bibinfo{person}{Jiaxin Chang}, \bibinfo{person}{Weinan Zhang}, {and} \bibinfo{person}{Linpeng Huang}.} \bibinfo{year}{2026}\natexlab{a}.
\newblock \showarticletitle{IssueExec: A Test-Driven Approach for Localizing Software Engineering Issues}.
\newblock \bibinfo{journal}{\emph{arXiv preprint arXiv:2607.17286}} (\bibinfo{year}{2026}).
\newblock


\bibitem[Liu et~al\mbox{.}(2026b)]%
        {liu2026GraphLocator}
\bibfield{author}{\bibinfo{person}{Wei Liu}, \bibinfo{person}{Chao Peng}, \bibinfo{person}{Pengfei Gao}, \bibinfo{person}{Aofan Liu}, \bibinfo{person}{Wei Zhang}, \bibinfo{person}{Haiyan Zhao}, {and} \bibinfo{person}{Zhi Jin}.} \bibinfo{year}{2026}\natexlab{b}.
\newblock \showarticletitle{GraphLocator: Graph-Guided Causal Reasoning for Issue Localization}.
\newblock \bibinfo{journal}{\emph{Proc. ACM Softw. Eng.}} \bibinfo{volume}{3}, \bibinfo{number}{FSE}, Article \bibinfo{articleno}{FSE051} (\bibinfo{date}{June} \bibinfo{year}{2026}), \bibinfo{numpages}{22}~pages.
\newblock
\href{https://doi.org/10.1145/3797079}{doi:\nolinkurl{10.1145/3797079}}


\bibitem[Ma et~al\mbox{.}(2026a)]%
        {ma2026llmagentscoderepositories}
\bibfield{author}{\bibinfo{person}{Dongjian Ma}, \bibinfo{person}{Silin Chen}, \bibinfo{person}{Yufei Yang}, \bibinfo{person}{Yuling Shi}, \bibinfo{person}{Yanfu Yan}, {and} \bibinfo{person}{Xiaodong Gu}.} \bibinfo{year}{2026}\natexlab{a}.
\newblock \bibinfo{title}{LLM Agents Can See Code Repositories}.
\newblock
\showeprint[arxiv]{2606.14061}~[cs.SE]
\urldef\tempurl%
\url{https://arxiv.org/abs/2606.14061}
\showURL{%
\tempurl}


\bibitem[Ma et~al\mbox{.}(2026b)]%
        {ma2026enhancing}
\bibfield{author}{\bibinfo{person}{Zexiong Ma}, \bibinfo{person}{Chao Peng}, \bibinfo{person}{Qunhong Zeng}, \bibinfo{person}{Pengfei Gao}, \bibinfo{person}{Yanzhen Zou}, {and} \bibinfo{person}{Bing Xie}.} \bibinfo{year}{2026}\natexlab{b}.
\newblock \showarticletitle{Enhancing Issue Localization Agent with Tool-Interactive Training}. In \bibinfo{booktitle}{\emph{Proceedings of the 2026 IEEE/ACM 48th International Conference on Software Engineering}} \emph{(\bibinfo{series}{ICSE '26})}. \bibinfo{publisher}{Association for Computing Machinery}, \bibinfo{address}{New York, NY, USA}, \bibinfo{pages}{692–703}.
\newblock
\showISBNx{9798400720253}
\href{https://doi.org/10.1145/3744916.3773137}{doi:\nolinkurl{10.1145/3744916.3773137}}


\bibitem[Meng et~al\mbox{.}(2026)]%
        {meng2026bugfix}
\bibfield{author}{\bibinfo{person}{Xiangxin Meng}, \bibinfo{person}{Zexiong Ma}, \bibinfo{person}{Pengfei Gao}, {and} \bibinfo{person}{Chao Peng}.} \bibinfo{year}{2026}\natexlab{}.
\newblock \showarticletitle{LLM-based Agents for Automated Bug Fixing: How Far Are We?}. In \bibinfo{booktitle}{\emph{Proceedings of the 2026 IEEE/ACM 48th International Conference on Software Engineering}} \emph{(\bibinfo{series}{ICSE '26})}. \bibinfo{publisher}{Association for Computing Machinery}, \bibinfo{address}{New York, NY, USA}, \bibinfo{pages}{855–867}.
\newblock
\showISBNx{9798400720253}
\href{https://doi.org/10.1145/3744916.3773269}{doi:\nolinkurl{10.1145/3744916.3773269}}


\bibitem[{OpenAI}(2026a)]%
        {openai2026codex}
\bibfield{author}{\bibinfo{person}{{OpenAI}}.} \bibinfo{year}{2026}\natexlab{a}.
\newblock \bibinfo{title}{Agents API: Build Durable Cloud Agents with a Managed Codex Harness}.
\newblock \bibinfo{howpublished}{OpenAI Developer Documentation}.
\newblock
\shownote{Accessed: 2026-09-23}.
\newblock
\urldef\tempurl%
\url{https://developers.openai.com/api/docs/guides/agents-api/overview}
\showURL{%
\tempurl}


\bibitem[{OpenAI}(2026b)]%
        {openai2026gpt56luna}
\bibfield{author}{\bibinfo{person}{{OpenAI}}.} \bibinfo{year}{2026}\natexlab{b}.
\newblock \bibinfo{title}{{GPT-5.6 Luna}}.
\newblock \bibinfo{howpublished}{\url{https://developers.openai.com/api/docs/models/gpt-5.6-luna}}.
\newblock
\shownote{Accessed: 2026-08-29}.
\newblock


\bibitem[{OpenCode}(2026)]%
        {opencode2026compaction}
\bibfield{author}{\bibinfo{person}{{OpenCode}}.} \bibinfo{year}{2026}\natexlab{}.
\newblock \bibinfo{title}{Compaction}.
\newblock \bibinfo{howpublished}{OpenCode Documentation}.
\newblock
\shownote{Accessed: 2026-09-23}.
\newblock
\urldef\tempurl%
\url{https://opencode.ai/v2/docs/compaction}
\showURL{%
\tempurl}


\bibitem[Pan et~al\mbox{.}(2024)]%
        {pan2024llmlingua2}
\bibfield{author}{\bibinfo{person}{Zhuoshi Pan}, \bibinfo{person}{Qianhui Wu}, \bibinfo{person}{Huiqiang Jiang}, \bibinfo{person}{Menglin Xia}, \bibinfo{person}{Xufang Luo}, \bibinfo{person}{Jue Zhang}, \bibinfo{person}{Qingwei Lin}, \bibinfo{person}{Victor R{\"u}hle}, \bibinfo{person}{Yuqing Yang}, \bibinfo{person}{Chin-Yew Lin}, \bibinfo{person}{H.~Vicky Zhao}, \bibinfo{person}{Lili Qiu}, {and} \bibinfo{person}{Dongmei Zhang}.} \bibinfo{year}{2024}\natexlab{}.
\newblock \showarticletitle{{LLML}ingua-2: Data Distillation for Efficient and Faithful Task-Agnostic Prompt Compression}. In \bibinfo{booktitle}{\emph{Findings of the Association for Computational Linguistics: ACL 2024}}, \bibfield{editor}{\bibinfo{person}{Lun-Wei Ku}, \bibinfo{person}{Andre Martins}, {and} \bibinfo{person}{Vivek Srikumar}} (Eds.). \bibinfo{publisher}{Association for Computational Linguistics}, \bibinfo{address}{Bangkok, Thailand}, \bibinfo{pages}{963--981}.
\newblock
\href{https://doi.org/10.18653/v1/2024.findings-acl.57}{doi:\nolinkurl{10.18653/v1/2024.findings-acl.57}}


\bibitem[Shi et~al\mbox{.}(2025)]%
        {shi2025longcodezip}
\bibfield{author}{\bibinfo{person}{Yuling Shi}, \bibinfo{person}{Yichun Qian}, \bibinfo{person}{Hongyu Zhang}, \bibinfo{person}{Beijun Shen}, {and} \bibinfo{person}{Xiaodong Gu}.} \bibinfo{year}{2025}\natexlab{}.
\newblock \showarticletitle{LongCodeZip: Compress Long Context for Code Language Models}. In \bibinfo{booktitle}{\emph{2025 40th IEEE/ACM International Conference on Automated Software Engineering (ASE)}} (Seoul, Korea, Republic of). \bibinfo{publisher}{IEEE Press}, \bibinfo{pages}{141–153}.
\newblock
\href{https://doi.org/10.1109/ASE63991.2025.00020}{doi:\nolinkurl{10.1109/ASE63991.2025.00020}}


\bibitem[Shi et~al\mbox{.}(2026)]%
        {shi2026codeocr}
\bibfield{author}{\bibinfo{person}{Yuling Shi}, \bibinfo{person}{Chaoxiang Xie}, \bibinfo{person}{Zhensu Sun}, \bibinfo{person}{Yeheng Chen}, \bibinfo{person}{Chenxu Zhang}, \bibinfo{person}{Longfei Yun}, \bibinfo{person}{Chengcheng Wan}, \bibinfo{person}{Hongyu Zhang}, \bibinfo{person}{David Lo}, {and} \bibinfo{person}{Xiaodong Gu}.} \bibinfo{year}{2026}\natexlab{}.
\newblock \showarticletitle{Codeocr: On the effectiveness of vision language models in code understanding}.
\newblock \bibinfo{journal}{\emph{arXiv preprint arXiv:2602.01785}} (\bibinfo{year}{2026}).
\newblock


\bibitem[Tian et~al\mbox{.}(2026)]%
        {zhao2026ensemble}
\bibfield{author}{\bibinfo{person}{Zhao Tian}, \bibinfo{person}{Pengfei Gao}, \bibinfo{person}{Junjie Chen}, {and} \bibinfo{person}{Chao Peng}.} \bibinfo{year}{2026}\natexlab{}.
\newblock \showarticletitle{Agent-Based Ensemble Reasoning for Repository-Level Issue Resolution}. In \bibinfo{booktitle}{\emph{Proceedings of the 2026 IEEE/ACM 48th International Conference on Software Engineering}} \emph{(\bibinfo{series}{ICSE '26})}. \bibinfo{publisher}{Association for Computing Machinery}, \bibinfo{address}{New York, NY, USA}, \bibinfo{pages}{1406–1418}.
\newblock
\showISBNx{9798400720253}
\href{https://doi.org/10.1145/3744916.3787761}{doi:\nolinkurl{10.1145/3744916.3787761}}


\bibitem[Wang et~al\mbox{.}(2024)]%
        {wang2024openhands}
\bibfield{author}{\bibinfo{person}{Xingyao Wang}, \bibinfo{person}{Boxuan Li}, \bibinfo{person}{Yufan Song}, \bibinfo{person}{Frank~F Xu}, \bibinfo{person}{Xiangru Tang}, \bibinfo{person}{Mingchen Zhuge}, \bibinfo{person}{Jiayi Pan}, \bibinfo{person}{Yueqi Song}, \bibinfo{person}{Bowen Li}, \bibinfo{person}{Jaskirat Singh}, {et~al\mbox{.}}} \bibinfo{year}{2024}\natexlab{}.
\newblock \showarticletitle{Openhands: An open platform for ai software developers as generalist agents}.
\newblock \bibinfo{journal}{\emph{arXiv preprint arXiv:2407.16741}} (\bibinfo{year}{2024}).
\newblock


\bibitem[Wang et~al\mbox{.}(2026)]%
        {wang2026swepruner}
\bibfield{author}{\bibinfo{person}{Yuhang Wang}, \bibinfo{person}{Yuling Shi}, \bibinfo{person}{Mo Yang}, \bibinfo{person}{Rongrui Zhang}, \bibinfo{person}{Shilin He}, \bibinfo{person}{Heng Lian}, \bibinfo{person}{Yuting Chen}, \bibinfo{person}{Siyu Ye}, \bibinfo{person}{Kai Cai}, {and} \bibinfo{person}{Xiaodong Gu}.} \bibinfo{year}{2026}\natexlab{}.
\newblock \showarticletitle{Swe-pruner: Self-adaptive context pruning for coding agents}.
\newblock \bibinfo{journal}{\emph{arXiv preprint arXiv:2601.16746}} (\bibinfo{year}{2026}).
\newblock


\bibitem[Wei et~al\mbox{.}(2025)]%
        {wei2025deepseekocr}
\bibfield{author}{\bibinfo{person}{Haoran Wei}, \bibinfo{person}{Yaofeng Sun}, {and} \bibinfo{person}{Yukun Li}.} \bibinfo{year}{2025}\natexlab{}.
\newblock \showarticletitle{Deepseek-ocr: Contexts optical compression}.
\newblock \bibinfo{journal}{\emph{arXiv preprint arXiv:2510.18234}} (\bibinfo{year}{2025}).
\newblock


\bibitem[Xia et~al\mbox{.}(2025)]%
        {xia2024agentless}
\bibfield{author}{\bibinfo{person}{Chunqiu~Steven Xia}, \bibinfo{person}{Yinlin Deng}, \bibinfo{person}{Soren Dunn}, {and} \bibinfo{person}{Lingming Zhang}.} \bibinfo{year}{2025}\natexlab{}.
\newblock \showarticletitle{Demystifying LLM-Based Software Engineering Agents}.
\newblock \bibinfo{journal}{\emph{Proc. ACM Softw. Eng.}} \bibinfo{volume}{2}, \bibinfo{number}{FSE}, Article \bibinfo{articleno}{FSE037} (\bibinfo{date}{June} \bibinfo{year}{2025}), \bibinfo{numpages}{24}~pages.
\newblock
\href{https://doi.org/10.1145/3715754}{doi:\nolinkurl{10.1145/3715754}}


\bibitem[Xiao et~al\mbox{.}(2026)]%
        {xiao2026agentdiet}
\bibfield{author}{\bibinfo{person}{Yuan-An Xiao}, \bibinfo{person}{Pengfei Gao}, \bibinfo{person}{Chao Peng}, {and} \bibinfo{person}{Yingfei Xiong}.} \bibinfo{year}{2026}\natexlab{}.
\newblock \showarticletitle{Reducing Cost of LLM Agents with Trajectory Reduction}.
\newblock \bibinfo{journal}{\emph{Proc. ACM Softw. Eng.}} \bibinfo{volume}{3}, \bibinfo{number}{FSE}, Article \bibinfo{articleno}{FSE056} (\bibinfo{date}{June} \bibinfo{year}{2026}), \bibinfo{numpages}{23}~pages.
\newblock
\href{https://doi.org/10.1145/3797084}{doi:\nolinkurl{10.1145/3797084}}


\bibitem[Yang et~al\mbox{.}(2024)]%
        {yang2024sweagent}
\bibfield{author}{\bibinfo{person}{John Yang}, \bibinfo{person}{Carlos~E Jimenez}, \bibinfo{person}{Alexander Wettig}, \bibinfo{person}{Kilian Lieret}, \bibinfo{person}{Shunyu Yao}, \bibinfo{person}{Karthik Narasimhan}, {and} \bibinfo{person}{Ofir Press}.} \bibinfo{year}{2024}\natexlab{}.
\newblock \showarticletitle{Swe-agent: Agent-computer interfaces enable automated software engineering}.
\newblock \bibinfo{journal}{\emph{Advances in Neural Information Processing Systems (NeurIPS 2024)}}  \bibinfo{volume}{37} (\bibinfo{year}{2024}), \bibinfo{pages}{50528--50652}.
\newblock


\bibitem[Zhang et~al\mbox{.}(2025)]%
        {zhang2025qwen3embedding}
\bibfield{author}{\bibinfo{person}{Yanzhao Zhang}, \bibinfo{person}{Mingxin Li}, \bibinfo{person}{Dingkun Long}, \bibinfo{person}{Xin Zhang}, \bibinfo{person}{Huan Lin}, \bibinfo{person}{Baosong Yang}, \bibinfo{person}{Pengjun Xie}, \bibinfo{person}{An Yang}, \bibinfo{person}{Dayiheng Liu}, \bibinfo{person}{Junyang Lin}, {et~al\mbox{.}}} \bibinfo{year}{2025}\natexlab{}.
\newblock \showarticletitle{Qwen3 embedding: Advancing text embedding and reranking through foundation models}.
\newblock \bibinfo{journal}{\emph{arXiv preprint arXiv:2506.05176}} (\bibinfo{year}{2025}).
\newblock


\bibitem[Zhang et~al\mbox{.}(2024)]%
        {zhang2024autocoderover}
\bibfield{author}{\bibinfo{person}{Yuntong Zhang}, \bibinfo{person}{Haifeng Ruan}, \bibinfo{person}{Zhiyu Fan}, {and} \bibinfo{person}{Abhik Roychoudhury}.} \bibinfo{year}{2024}\natexlab{}.
\newblock \showarticletitle{AutoCodeRover: Autonomous Program Improvement}. In \bibinfo{booktitle}{\emph{Proceedings of the 33rd ACM SIGSOFT International Symposium on Software Testing and Analysis}} (Vienna, Austria) \emph{(\bibinfo{series}{ISSTA 2024})}. \bibinfo{publisher}{Association for Computing Machinery}, \bibinfo{address}{New York, NY, USA}, \bibinfo{pages}{1592–1604}.
\newblock
\showISBNx{9798400706127}
\href{https://doi.org/10.1145/3650212.3680384}{doi:\nolinkurl{10.1145/3650212.3680384}}


\end{thebibliography}

\end{document}